\documentclass[3p]{elsarticle}

\usepackage{fancyhdr}
\usepackage{booktabs}
\usepackage{ctable}
\usepackage{amsmath}
\usepackage{mathtools}
\usepackage{commath}
\usepackage{graphicx}
\usepackage{nomencl}
\usepackage{commath}
\usepackage{natbib}
\usepackage{tabularx}
\usepackage{longtable}

\usepackage{multirow}
\usepackage{longtable}
\usepackage{textcomp}
\usepackage{array}
\usepackage{graphics}
\usepackage{tabularx}
\usepackage{color}
\usepackage{geometry}
\usepackage{natbib}
\usepackage{mciteplus}
\usepackage{footnote}
\usepackage{amssymb}
\usepackage{pdflscape}
\usepackage{afterpage}
\usepackage{multirow}
\usepackage[title]{appendix}
\usepackage{subfigure}
\usepackage{caption}
\usepackage{multirow}
\usepackage{amssymb}
\usepackage{amsmath}
\usepackage{mathtools}
\usepackage{ragged2e}
\usepackage{blindtext}
\usepackage{commath}
\usepackage{amssymb}
\usepackage{natbib}
\usepackage{mleftright} 
\usepackage{latexsym}
\usepackage{color,soul}

\usepackage{etoolbox}
\renewcommand\nomgroup[1]{%
	\item[\bfseries
	\ifstrequal{#1}{A}{}{%
		\ifstrequal{#1}{G}{Greek Letters}{%
			\ifstrequal{#1}{S}{Subscripts}{}}}%
	]}

\usepackage[version=3]{mhchem} 

\journal{Industrial \& Engineering Chemistry Research}

\biboptions{numbers,sort&compress}

\begin{document}
\begin{frontmatter}

\title{Numerical insights from population balance model into the distribution of bitumen residues in industrial horizontal pipes during the hydrotransport  of oil sands tailings}

\author[1,2]{Somasekhara Goud Sontti\corref{cor1}}
\ead{somasekhar.sonti@iitdh.ac.in}
\cortext[cor1]{Corresponding author}
\address[1]{Department of Chemical and Materials Engineering, University of Alberta, Alberta T6G 1H9, Canada}
\address[2]{Multiphase Flow and Microfluidics (MFM) Laboratory, Department of Chemical Engineering, Indian Institute of Technology Dharwad, Dharwad, 580011, Karnataka, India}

\author[1]{Xuehua Zhang}

\newpage
\begin{abstract}

Pipeline transportation is a vital method for conveying crushed oil sand ores and tailings in the oil sands industry. This study focuses on enhancing economic benefits by exploring the separation of valuable bitumen residues from coarse sand tailings within hydrotransport pipelines. Employing three-dimensional transient Eulerian-Eulerian computational fluid dynamics (CFD) simulations coupled with a population balance model (PBM), we examine the aggregation and breakage of bitumen droplets under various flow conditions. The CFD-PBM model's accuracy is validated against field measurements of velocity profiles and pressure drops. Our findings reveal that higher slurry velocities lead to intensified particle-bitumen interactions, resulting in reduced aggregated bitumen droplet sizes at the pipeline's core. Additionally, variations in bitumen fraction cause shifts in the distribution of coarse particles along the pipe's vertical axis, with increased aggregation and larger droplets in the upper region. Notably, we demonstrate that smaller bubbles promote a more uniform distribution of bitumen compared to larger bubbles. These insights provide valuable knowledge for optimizing bitumen recovery processes, facilitating the integration of pipeline dynamics with downstream separation and extraction units.
	
\end{abstract}


\end{frontmatter}

\newpage
\section{Introduction} 
Concentrated slurry transported through long pipelines has been  widely used in various industrial applications, such as coal slurries, wastewater treatment, mineral extraction, transport of ores, and mining tailings~\citep{pullum2018hydraulic,plumlee2011mine,mohaibes2004aerobic,scoble2003mining}. Oil sand ore and tailings are commonly transported in horizontal pipelines transport due to its economic advantage over railroad transportation and much-reduced noise disturbance to the environment, particularly when mines are located in remote areas. Concentrated slurries are multi-component and multiphase systems that include solid particles, water, bubbles, oil droplets, and other substances.\\

Transport of concentrated oil sands tailings in horizontal pipelines is challenging due to the complex interaction between the particle and bitumen-droplet collision and friction between the multi-sized particles and the walls in a highly turbulent flow. Simulating non-Newtonian slurry systems is also challenging due to fluid turbulence. Oil sand tailings consist of a tiny fraction of bitumen residue (0.1 to 3 wt \%) and a high solid content that is transported to tailing ponds. Although the bitumen residue is very low in concentration, it is desirable to separate and recovery the bitumen fraction from tailings, due to the vast volume of the tailings,  for economic gain and additional environmental benefits~\citep{khademi2018provenance,dibike2018modelling}. Tailings flow and bitumen separation from tailings before disposal into tailings ponds must be considered when designing a separation technology. The complex interactions between multiple secondary phases (i.e., dispersed phases) pose a significant challenge to slurry flow characteristics in pipelines~\citep{ruan2021effect}. 


To date, numerous researchers have conducted slurry flow experimental studies to estimate flow regimes and pressure drop in horizontal pipelines. Interestingly, microbubble injection is found to be effective at enhancing bitumen recovery. Bubbles with small sizes (such as nanobubbles) can lead to the attachment of large bubbles onto solid particle surfaces, and possibly improve the separation efficiency ~\citep{scales2011,zhang2014}.
Experimental studies have demonstrated that microbubbles in the slurry flow improve the recovery efficiency of bitumen in a lab-scale pipeline~\citep{motamed2020microbubble,zhou2022microbubble,zhou2023}.\\

\noindent The earlier reported CFD model results have improved our understanding of tailings transport in pipelines and underlying factors that impact bitumen distribution and recovery. Recent work of Sadeghi et al.~\cite{sadeghi2023computational} numerically simulated a complex slurry system using a mixture multiphase model considering 8 different size solid particles, bitumen, gas bubbles and a carrier liquid. A comprehensive examination was conducted to evaluate the influence of mixture velocity, particle size distribution, carrier fluid rheology, and pipe angle on the performance of the slurry flow. The key aspects of the study were to assess the variations in bitumen concentration, solids concentration and pressure drop. For complex multiphase flows, the Eulerian-Eulerian (E-E) two-fluid model (TFM) is a powerful way to predict hydrodynamic properties~\citep{messa2021computational,parvathaneni2021eulerian} and also offers superior computational efficiency. In our previous work Sontti et al.~\cite{sontti2022computational}, bitumen residue transport in the industrial-scale horizontal pipeline were simulated by using E-E multi-fluid model where bitumen droplets are considered to be of a constant size in the tailings transport.\\



\noindent The Computational Fluid Dynamics-Population Balance Model (CFD-PBM) offers a highly promising solution for a comprehensive analysis of bitumen droplet agglomeration and breakage frequency. This model integrates the principles of CFD and PBM, allowing for a thorough and precise examination of the intricacies of fluid systems. The purpose of CFD-PBM is to provide a more accurate and comprehensive understanding of fluid and particle behavior in various applications. Cai et al.~\cite{cai2019investigation} developed a model to represent an ice slurry pipeline transport in an elbow pipe by a CFD-PBM approach based on E-E TFM. The CFD-PBM model included an evaluation of both particle aggregation and fragmentation. The outcome indicated that the diameter of ice increases significantly due to aggregation as it moves along the flow direction. The effect of the change in  ice packing fraction and slurry velocity was reported on concentration profiles and particle size.\\

\noindent Recently, ~\citet{ma2022numerical} also studied a similar system using the CFD-PBM coupled model to investigate ice slurry flow characteristics. According to their findings, elbow angles greatly affect pressure, velocity field, ice particle size, and distribution of ice slurry. The results indicate that aggregation takes place at the pipe section center at elbow outlet sections where the elbow angle is less than 90°. A multiphase flow model that considers bubble coalescence and hydrate breakage has been developed by Cao et al.~\cite{cao2022gas} for studying gas-liquid-hydrate slurry flow in a vertical pipe. The population balance theory described the bubble and hydrate particle size evolution. According to their findings, bubbles are likely to aggregate at the center of the pipe during the flow process due to inhomogeneous phase distribution. The work by Balakin et al.~\cite{balakin2016modelling} considered non-Newtonian  rheology and developed a CFD-PBM model for predicting the details of hydrate phase formation, agglomeration of solids, and granular interactions in industrial pipelines. However, they clearly stated that the choice of a non-Newtonian model is unclear. \\



\noindent Considering the advantage of the CFD-PBM approach, where the bitumen droplet sizes are evaluated indirectly at every cell/time step and can benefit from a source of additional information on secondary phase distribution in the computational domain. 
We first demonstrate the CFD-PBM model's capability of predicting bitumen residues in oil sand tailings transport, which has never been reported before. The 3D CFD-PBM model was further utilized to evaluate the distribution of bitumen droplets under different flow conditions in the tailings system. The outcomes of this research offer a comprehensive understanding of bitumen droplet distribution and can be applied in the design of pipelines and recovery processes for bitumen. \\



\section{Field data of velocity and pressure drop} 
\begin{figure}
	\centering
	\includegraphics[width=0.6\textwidth]{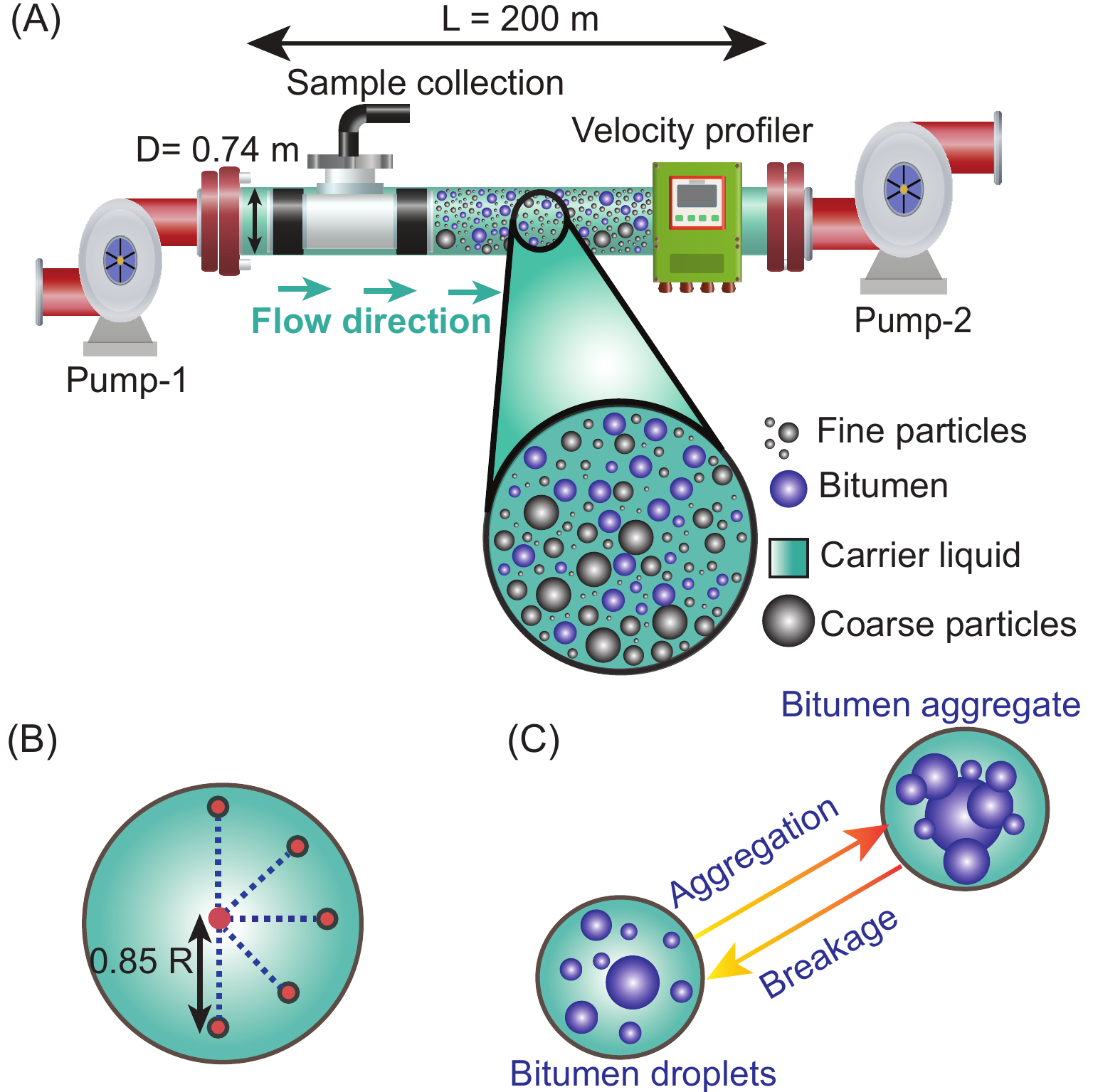}
	\caption{\label{fig:Loop} (A) A schematic illustration of the tailings hydrotransport pipeline, including its dimensions. The magnified view displays the typical composition of the tailings, consisting of coarse and fine particles along with bitumen droplets in a fluid carrier. (B) depicts velocity profiler measurements at five different points, and (C) shows a schematic representation of the formation and breakage of bitumen droplet aggregates into smaller droplets.}
\end{figure}

\noindent A schematic of a horizontal industrial pipeline employed to collect field data is shown in Fig.\ref{fig:Loop}. An inlet and an outlet of the pipeline are connected by two pumps. The length of the pipeline is 200\,m, and the diameter is 74\,cm. To validate the CFD model, a significant amount of field data is gathered from a section of an industrial pipeline in an oil sand tailings system. All samples are collected every twelve hours in the center of the pipe at the first pump discharge, as shown in Fig.\ref{fig:Loop}a. After that, a Dean\textendash Stark setup is used to evaluate the composition of the mixture~\cite{bulmer1979syncrude}. A sieving method procedure is employed to quantify the particle size distribution (PSD) for solid particles. \\

A non-invasive SANDtrac velocity profile system (CiDRA) is mounted to the pipeline near the outlet to measure velocity distribution across the pipeline. The velocity profiler is an array of five sensors attached to the pipe wall, which track the turbulent eddies inducing pressure disturbance and, consequently, forces on the wall. An array of sensors converts the forces acting on sensors to electrical signals, and the velocity value is calculated from these signals~\cite{maron2008new}. This technique is well–established and applied commercially for
slurry transport in the industry, where detection of the formation of stationary beds is a usage of the
device.\\

The velocity field data is collected every two seconds at five points, as shown in Fig.\ref{fig:Loop}b. The system collected millions of data points systematically analyzed, and the fluctuations are also noticed from the velocity points, which are recorded from the hydrotransport pipeline. From the velocity profiler plot, we have considered a 30-minute time window by considering the minimum standard deviation of lower than 0.2. The corresponding velocity is considered as a flow velocity for the specific time window~\cite{sontti2022computational}.  Additionally, two gauge pressure meters are used to measure pressures at the discharge and suction of the first and second pumps, respectively. The pressure gradient is computed by utilizing the average values of pressure at the first pump's discharge and the second pump's suction. In the end, a set of cases are prepared based on the field data to validate the CFD model. A detailed explanation of the mixture composition and parameters for model validation is available in our previous work Sontti et al.~ (\cite{sontti2022computational};~refer to Table I).

\section{Modeling of multiphase flows} 

Simulating slurry flows with CFD (Computational Fluid Dynamics) can be accomplished through two main methodologies: Eulerian-Eulerian (E-E) and Eulerian-Lagrangian (E-L) \citep{messa2021computational}. The E-E method considers both the fluid and solid phases as continuous and homogeneous and models their interactions through the Navier-Stokes equations and turbulence models. On the other hand, the E-L method treats the fluid phase as continuous and the solid phase as discrete particles, tracked using particle tracking algorithms. The KTGF (Kinetic Theory of Granular Flow) can be utilized in conjunction with the E-E model to account for the impact of solid particle interactions on the flow field~\cite{parvathaneni2023eulerian,parvathaneni2023effect}. The choice of methodology and coupling with KTGF will depend on the simulation's specific needs and the desired accuracy level. In contrast, the E-L model tracks each secondary phase, such as solid particles, droplets, and bubbles, individually using Newton's law of motion. Although the E-L model is more computationally intensive than the E-E model, it is applicable to dilute systems \citep{messa2020analysis,wang2022multi,ALOBAID2022}.\\

The effects of different particle sizes on flow characteristics must be considered. The Eulerian multiphase model is an appropriate strategy for investigating high-concentration slurry transport in pipelines, notably that involving multiple particles and droplets. In light of this, in the current study, we used the FVM based on E\textendash E multi\textendash fluid model (MFM) to study the tailings residuals flow characteristics.  

\subsection{Eulerian-Eulerian multiphase flow model}
The Eulerian multiphase model considers phases as interpenetrating continuous systems. Each phase is solved individually for the continuity and momentum equations. Mass, momentum, and energy conservation equations should be satisfied by all phases. In the present CFD modeling, the slurry flow system contains four phases. The continuous phase is sand particle suspensions in water with concentrations ranging from 10\% to 40\%. The carrier liquid with fine particles is considered as a Casson rheological model \citep{rheology}.  The solids (i.e., solid-1 and solid-2), bitumen droplets, and gas bubbles are treated as dispersed phase. Solid phases are solved similarly to the fluid phase in the Eulerian framework. Table~\ref{tab:CFD_model_equations_Binary} lists the governing equations for mass and momentum of both primary and secondary phases. Equation (1) represents the continuity equation for all the phases (i.e., carrier, bitumen, solids) considered, and Eq. (2) represents the momentum equation for the carrier fluid. In Equation (2), $\vec{F}_{\mathrm{l},s}$ and $\vec{F}_{\mathrm{l},b}$  represent the forces between the liquid and solid and the liquid and bitumen, respectively~\cite{sontti2022computational}. Equations 3 and 4 represent the moment equations for solid and bitumen/bubble phases.\\


\begin{table}[htbp]
	\caption{Momentum equations.~\citep{fluent2011ansys, sadeghi2022cfd}}
	\label{tab:CFD_model_equations_Binary}
	\centering
	\renewcommand{\arraystretch}{2}
	\begin{tabular}{p{3cm} p{13cm}}
		\hline
		\textbf{Equation} & \textbf{Description} \\ [0.5ex] 
		\hline
		Continuity & $\frac{\partial}{\partial t}(\rho_q\alpha_q) + \nabla.(\rho_q\alpha_q\mathbf{v}_q) = 0$, \hspace{0.5cm} $(q=l,s,b)$~~ (1) \\
		\hline
		Momentum (Carrier) & $\frac{\partial}{\partial t}(\rho_l\alpha_l\mathbf{v}_l) + \nabla.(\rho_l\alpha_l\mathbf{v}_l\mathbf{v}_l) = -\alpha_l\nabla P + \rho_l\alpha_l\mathbf{g} + \nabla .\tau_l + \mathbf{F}_{1,s} + \mathbf{F}_{1,c} + \mathbf{F}_{1,b}$~~(2) \\
		\hline
		Momentum (Solid Phase) & $\frac{\partial}{\partial t}(\rho_{si}\alpha_{si}\mathbf{v}_{si}) + \nabla.(\rho_{si}\alpha_{si}\mathbf{v}_{si}\mathbf{v}_{si}) = -\alpha_{si}\nabla P - \nabla P_{si} + \rho_{l}\alpha_{si}\mathbf{g} + \nabla .\tau_{si} + \mathbf{F}_{si,l} + \mathbf{F}_{\text{drag},si,c} + \mathbf{F}_{\text{drag},si,b}\beta_{ij} (\mathbf{v}_{sj} - \mathbf{v}_{si})$~~(3)
		) \\
		\hline
		Momentum (Bitumen or Bubble) & $\frac{\partial}{\partial t}(\rho_{b}\alpha_{b}\mathbf{v}_{b}) + \nabla.(\rho_{b}\alpha_{b}\mathbf{v}_{b}\mathbf{v}_{b}) = -\alpha_{b}\nabla P + \rho_{l}\alpha_{b}\mathbf{g} + \nabla .\mu_{b}(\nabla\mathbf{V} + \nabla\mathbf{V}^T) + \mathbf{F}_{b,l} + \mathbf{F}_{b,si} + \mathbf{F}_{b,ci}$~~ (4) \\
		\hline
	\end{tabular}
\end{table}


\vspace{0.6cm}

\noindent The model accounts for the liquid-solid interactions through the inclusion of the drag force and the turbulent dispersion force. In addition, the interactions between the carrier fluid and bitumen droplets are also considered in the model. Secondary phase solid particle and bitumen momentum equations are also expressed in the CFD-PBM model. Particle interactions are described by the KTGF model. Furthermore, the constitutive equations derived from granular flow theory are used to solve the stress-strain tensors of the solid phase. The constitutive equations in this work are identical to those in Sontti et al.~\cite{sontti2022computational} and are, therefore, not repeated in this section in explicit detail. The phase coupling in the model is regulated by the pressure and interphase exchange coefficients. The model also takes into account energy dissipation, particle energy exchange, and inter-phase forces. A drag model and constitutive equation should be used when solving these conservation equations. A list of the drag models considered in the CFD-PBM model and KTG model properties are listed in Table~S1. As a result of systematic comparisons between carrier-solid and carrier-bitumen drag models, Gidaspow et al.~\cite{gidaspow1991hyd} and Symmetric model~\citep{fluent2011ansys,sen2016cfd} drag models are used to capture interface force between carrier-solid and carrier-bitumen.


\subsection{Turbulence model}
Numerous studies have demonstrated that $k$-$\epsilon$ and $k$-$\omega$ turbulence models feasibility for different systems~\citep{li2018hydrodynamic,liu2022effect}. The standard $k$-$\epsilon$ model is suitable for modeling complex slurry flows. This study also utilizes a mixture turbulence model $k$-$\epsilon$ based on Reynolds-averaged Navier-Stokes (RANS) equations, similar to previous studies \citep{li2018hydrodynamic,puhan2023insights,liu2022effect}, to describe turbulent slurry flow. In our previous study, we have described the $k$-$\epsilon$ turbulence model governing equations in more detail (refer to Table III). In this regard, readers may refer to our earlier work of Sontti et al.~\cite{sontti2022computational} for additional information and solver settings. In addition, to achieve accurate information on solid concentrations and transport properties near walls, enhanced wall treatment (EWT) is used, which combines enhanced wall functions with two-layer models \citep{fluent2011ansys,li2018effect}. A recent study of Shi et al.~\cite{shi2020experimental} showed that the turbulence model with enhanced wall treatment for high solid concentration systems was more accurate at predicting slurry flows.


	\subsection{Non-Newtonian rheological model for carrier fluid }
	\noindent Sand particle suspensions in water with concentrations ranging from 10\% to 40\% are described by the Casson rheological model \citep{rheology}, and this range is similar to that in the current study. The Casson model provides a good fit to the viscosity-shear rate behavior of many yield stress fluids and is commonly used in industrial applications. In recent years, many researchers have effectively modeled the rheological behavior of tailings slurries using the Casson model \citep{britocasson,sontti2022computational}. Therefore, the Casson model is used to model the non-Newtonian behavior of the carrier liquid in this study since the mass fraction of fine particles falls within the specified range. The rheological equation for this model is given by Eq. \eqref{eqn:Cassonmodel}, where $\mu_c$ represents the Casson viscosity.
	
	
	\begin{equation}  \tag{5}\label{eqn:Cassonmodel}
		\tau_{{l}}^{1/2}=\tau_{{y}}^{1/2}+\mu_{{c}}^{1/2} \dot{\gamma}^{1/2}
	\end{equation}
	Where $\tau_l$ is shear stress, $\tau_{{y}}$ represents yield stress, and $\mu_c$ is Casson viscosity. The fluid's viscosity is then described as a function of both the yield stress and the shear rate. The rheological behavior of a continuous phase Eq. \eqref{eqn:Cassonmodel} is expressed as continuous phase shear stress $\tau_l$ and Eq. 5 plugged into Eq. (2) in the continuous phase momentum equation. A user-defined function (UDF) is applied to the CFD-PBM model to implement the carrier viscosity rheological model. 
	\subsection{Population balance model}
	\noindent The PBM is appropriate for dealing with secondary phases since it solves the number of secondary phases like solids/droplets  based on the population balance equation (PBE). In general, the PBE describes the conservation of the number density function of the bitumen phase to solve the PBM equation; a discrete method is used. A benefit of this method is that it provides the particle size distribution (PSD)  directly and has robust numerically. The CFD-PBM is an integrated method derived from Boltzmann statistical transport equation to simulate the multiphase flow. A list of PBM model equations based on Ramkrishna et al.~\cite{ramkrishna2000population} are presented in Table.\ref{tab:PBM}.\\

\begin{table}[!ht]
	\centering
	\caption{Population balance modelling equations.}
\label{tab:PBM}
\setcounter{equation}{5}
	\renewcommand{\arraystretch}{1.5}
	\begin{tabular}{l r}
			\hline
		\textbf{Equation} & \\
		$\frac{\partial n_i}{\partial t}+\nabla \cdot\left(\overrightarrow{u_b} n_i\right)=S_i$ & 6 \label{eq:PBE1} \\
		$S_i=B_{a, i}-D_{a, i}+B_{b, i}-D_{b, i}$ & 7 \\
		$B_{a, i}=\sum_{v_{i-1} \leqslant v_j+v_k \leqslant v_i}^{j \geqslant k}\left(1-\frac{1}{2} \delta_{j k}\right) \theta\left(v_j+v_k\right) a\left(v_{j,} v_k\right) n_j n_k$ & 8 \\
		$D_{a, i}=\sum_{j=1}^M a\left(v_i, v_j\right) n_i n_j$ & 9 \\
		$B_{b, i}=\sum_{j=1}^M v \theta_{i j} g\left(v_i\right) n_i$ & 10 \\
		$D_{b, i}=g\left(v_i\right) n_i$ & 11 \\
		$\theta=\begin{cases}
			\frac{v_{i+1}-\left(v_j+v_k\right)}{v_{i+1}-v_i}, & v_{i \leqslant}\left(v_j+v_k\right) \leqslant v_{i+1} \\
			\frac{\left(v_j+v_k\right)-v_{i-1}}{v_i-v_{i-1}}, & v_{i-1 \leqslant}\left(v_j+v_k\right) \leqslant v_i
		\end{cases}$ & 12 \\
		$\theta_{i j}=\int_{v_i}^{v_{i+1}} \frac{v_{i+1}-v}{v_{i+1}-v_i} \beta\left(v, v_j\right) d v+\int_{v_i}^{v_{i+1}} \frac{v-v_{i-1}}{v_i-v_{i-1}} \beta\left(v, v_j\right) d v$ & 13 \\
		\hline
	\end{tabular}
\end{table}

	The PBM solves a set of PBE's equations of time change rates of the number density of bitumen droplets. Bitumen droplet sizes are discretized into $n$ subclasses as expressed in Eq.\,\eqref{eq:PBE1}. The source term is denoted with $S_{i}$, and $i$th represents the droplet class formed by droplet breakage and agglomeration due, which is described in Eq.\,\eqref{eq:sourceterm}. The birth and death rates caused by agglomeration and breakup of $i$th are expressed from Eq.\,\eqref{eq:BA} to Eq.\,(13).  According to this method, the bitumen droplet breakage can be modeled based on the interaction between bitumen droplets and turbulent eddies, which leads to simulate the bitumen droplet deformation. The CFD-PBM model systematically evaluated the sensitivity of the number of bins while maintaining the minimal and maximum bitumen droplet sizes. A negligible change in pressure drop and velocity profile is observed. As a result, computation time is increased significantly. Taking this into consideration,  six bins are considered in the CFD-PBM model.\\
	
	The bitumen droplet diameters are discretized into six intervals ranging from 10 to 570 $\mu$m  in this study. A bubble coalescence and a breakup closure are required to account for mass and momentum transfer between droplet classes. The Luo and Svendsen~\cite{luo1996theoretical} model still provides mechanistic insights into the underlying physical processes of particle/droplet aggregation and breakage. It captures the key phenomena and trends observed in experiments, allowing researchers to understand the dominant mechanisms and identify critical factors affecting particle behavior. In the present study, we have considered the Luo and Svendsen~\cite{luo1996theoretical} model for both breakage and agglomeration, followed by previous works \citep{li2021cfd,thakur2022hydrodynamic,maluta2023experimental}. In practical cases, sand particles can also form agglomerates when bitumen acts as a binder. However, in our present CFD-PBM study, we mainly focused on the bitumen residual distribution in the pipeline. It is also important to note that the bitumen fraction is too low. Therefore, sand particle agglomeration caused by bitumen droplets is omitted.
	\section{Computational modeling}
	Pipeline transportation of slurries with high concentrations and multiple particle sizes can be modeled and analyzed with the Euler multiphase flow model. We investigated secondary phase droplet distribution in an industrial\textendash scale horizontal pipeline tailings system by using the CFD-PBM model. According to FVM formulation and considerations, volume-averaged continuity and momentum equations were solved. In this work, to solve all unsteady state equations, we use the FVM-based commercial software \textit{ANSYS Fluent} solver 2022 R2 and all the results are post-processed in CFD-post and Matlab, respectively. 
	
	\subsection{Geometry and boundary conditions}
	
	This study examines a large-scale circular pipeline, with a diameter of 0.74 meters and a length of 105 meters. It includes the inlet, outlet, and wall of the horizontal pipe. The volume fraction of each phase is specified at the inlet where the velocity of each phase is considered to be uniform. The outlet is subject to atmospheric pressure conditions. The no-slip condition is applied at the wall, resulting in a zero velocity for the liquid phase. Turbulence intensities and viscosities are set at 5 and 10 $(N. s/m^2)$  for all phases (i.e., carrier, solids, bitumen, and bubble), based on prior research by Li et al.~\cite{li2018hydrodynamic}. The rheological properties, parameters of the KTGF model, and PBM bin details are provided in Tables \ref{tab:Materialparameters} and \ref{tab:Bitumeninitialsize}.\\
	
	
	\begin{table}[!ht]
		\centering
		\caption{Fluid properties and boundary conditions used in CFD simulations.}
		\vspace{0.2cm}
		\label{tab:Materialparameters}
		\renewcommand{\arraystretch}{0.8}
		\resizebox{ 0.55\textwidth}{!}{%
			\begin{tabular}{ll}
				\specialrule{.1em}{.05em}{.05em} 
				\multicolumn{1}{c}{Parameter}                      & \multicolumn{1}{c}{Value} \\ \specialrule{.1em}{.05em}{.05em} 
				Pipe diameter \& length (m)                                 & 0.74 \& 105          \\
				Sand particle size ($\mu$m)                        & 75 \& 700           \\
				Sand particle density ($kg/m^{3}$)              & 2650           \\
				Carrier density (water+fines) ($kg/m^{3}$)                       & 1335           \\
				Bitumen viscosity at 20 \textdegree C ($Pa.s$)                         & 20           \\
				Casson  viscosity $\mu_{c}$ ($Pa^{1/2} s^{1/2}$)                         & 0.0035 \citep{rheology}          \\
				Yield stress, $\tau_{y}$ ($Pa$)                         & 0.0016  \citep{rheology}         \\
				Restitution coefficient   & 0.90 \citep{li2018hydrodynamic,zhang2021influence}        \\
				Specularity coefficient                                & 0.20 \citep{liu2021computational}          \\
				
				\specialrule{.1em}{.05em}{.05em} 
			\end{tabular}%
		}
	\end{table}
	
\begin{table}
	\centering
	\caption{Initial size distribution of bitumen droplets.}
	\vspace{0.1cm}
	\label{tab:Bitumeninitialsize}
	\renewcommand{\arraystretch}{0.60}
	\resizebox{0.7\textwidth}{!}{%
		\begin{tabular}{lcl}
			\toprule
			\text{Bin number} & \text{Mean diameter ($\mu$m)} & \text{Volume fraction} \\
			\midrule
			Bin-0 & 570 & 0 \\
			Bin-1 & 253 & 0 \\
			Bin-2 & 113 & 100\% \\
			Bin-3 & 50 & 0 \\
			Bin-4 & 22 & 0 \\
			Bin-5 & 10 & 0 \\
			\bottomrule
		\end{tabular}%
	}
\end{table}

	\subsection{Model implementation and solver settings}
	
	Interphase exchange coefficients are used to model secondary-phase interactions. The interphase exchange coefficients and pressure coefficients of all phases are coupled. A second-order implicit scheme is employed to discretize the transient formulation, and the convergence criterion of $10^{-4}$ for each scaled residual component is imposed between two successive iterations. The simulation is performed for a flow time of 200 s with a fixed time step of 0.01 s and a gravitational acceleration of $-9.8$ m$^2$/s considered in the Y-direction of the pipeline. The numerical results are quantified at Z-90 m along the YX plane, and distribution profiles are analyzed in the central region of the YX by plotting a vertical reference line. The simulation is performed using the high-performance computing facility at Compute Canada Graham cluster with 44 CPUs. Detailed information on the solver settings and schemes can be found in Table S1 (in Supporting Information).


	\subsection{ Mesh independence study}
	To check the influence of mesh density on CFD-PBM model results, we performed a set of simulations with three different mesh designs by altering the number of nodes. A 3D computationally structured mesh along the length of a pipe and its cross-section is depicted in Figure S1. Three mesh designs are examined to ensure reliable data: coarse, fine, and extra fine. A coarse mesh has 1,68,682 nodes, a fine mesh has 3,39,500 nodes, and an extra fine mesh has 5,27,253 nodes. Flow physics can be captured accurately with a fine grid and the corresponding number of nodes and the average mesh element size 2 mm by comparing velocity profiles for three different meshes \citep{sontti2022computational}. 
	
	The values for wall \(y^{+}\) are constantly monitored for the "fine" grid used for the simulations. The maximum \(y^{+}\) for the secondary phases (i.e., bitumen and solids) ranges from 0.068 to 80, and for the carrier phase is approximately 160, which clearly indicates that the grid resolution near the wall is fine enough. This is crucial for accurately capturing near-wall phenomena. As a reference, for high Reynolds number flows, it's a common guideline to maintain \(y^{+}\) values between 30 and 300. This range is considered suitable for a wide variety of industrial applications, especially when using the \(k\)-\(\epsilon\) turbulence model. These \(y^{+}\) values ensure that the flow near walls is appropriately resolved and contributes to accurate simulations. 
	\section{Model verification with industrial-scale field data}
\begin{figure}
		\centering
		\includegraphics[width=0.7\textwidth]{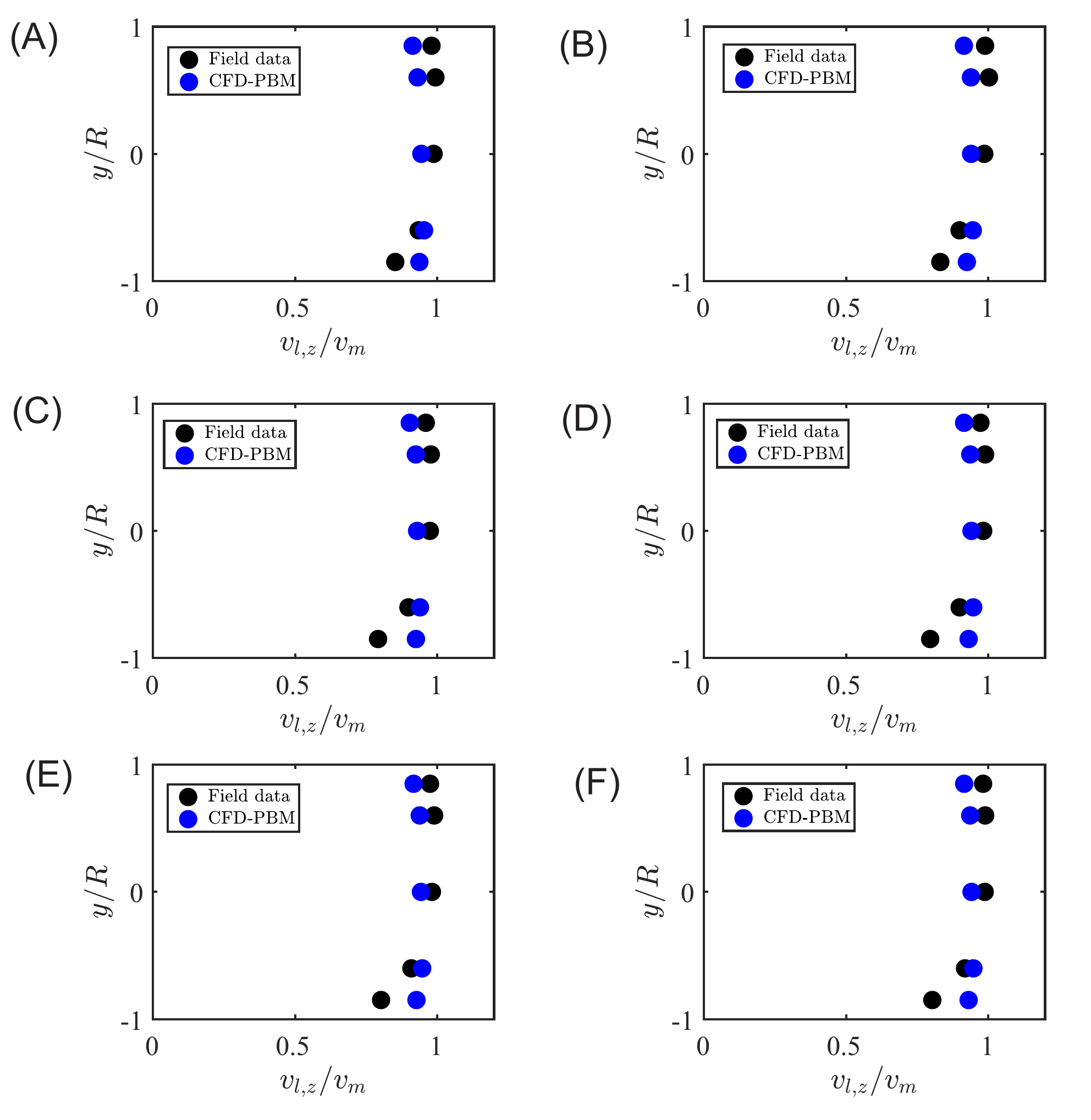}
		\caption{\label{fig:ValidationCFD} 
			The velocity distribution of field data is compared with the CFD-PBM model for six cases (A) to (F). The CFD-PBM predictions are represented by blue circles, and the field data is represented by black. The fluid properties and flow conditions details can be found in Sontti et al.~ \cite{sontti2022computational} in Table I. The tailings system field data is also reported in the same table.}
	\end{figure}

	Experimental data on bitumen droplet distribution within such complex multiphase flow in pipeline systems are limited. The existing field data on oil sands tailings hydrotransport in an industrial horizontal pipe supports our CFD-PBM model findings with pressure drop and velocity distribution. At first, we systematically demonstrated the efficacy of the developed CFD-PBM model in predicting reliable results for six field data sets as shown in Fig.\ref{fig:ValidationCFD}. The model prediction is compared mainly with velocity distribution and pressure gradient. According to the CFD model, the carrier fluid velocity field data are in excellent agreement with the model predictions. It is evident that the model forecasts with a maximum average velocity lower than 8\% as shown in Table.~\ref{tab:validation}.\\
	
	\noindent For all the six considered cases, an exceptional agreement was found at the middle of the pipe where maximum flow velocity can occur in the pipeline. However, the maximum deviation is observed near the bottom wall, where the coarse particles can accumulate or flow near the bottom wall of the pipeline in a sliding bed. Near the top wall also showed minimal deviation due to the accumulation of fine particles, bitumen droplets, and, importantly bitumen aggregates. It is important to remark that the velocity agreement deviation is attributable due to several factors, such as particle-wall interactions and near-the-wall modeling considerations and assumptions. On the other hand, precise measurement of the velocity field near the wall also poses difficulties in the slurry flow due to the static/dynamic bed formation. Nevertheless, the CFD-PBM model agreement showed excellent agreement with six-time windows, especially for an industrial-scale system with field data.  The CFD simulations, coupled with a population balance model (PBM) results, will provide insight into the effects of bitumen droplet aggregation and breakage on flow characteristics.\\ 
	
	\begin{table}
		\caption{Velocity and pressure drop error percentages for six field data cases.}
		\vspace{0.3cm}
		\label{tab:validation}
		\centering
		\renewcommand{\arraystretch}{0.60}
		\begin{tabular}{ccccc}
			\specialrule{.1em}{.05em}{.05em}
			Case & \begin{tabular}[c]{@{}c@{}}Average velocity \\ error \%\end{tabular} & \begin{tabular}[c]{@{}c@{}}Field data \\ $\Delta P/L$ (kPa/m)\end{tabular} & \begin{tabular}[c]{@{}c@{}}CFD-PBM prediction \\ $\Delta P/L$ (kPa/m)\end{tabular} & \begin{tabular}[c]{@{}c@{}}$\Delta P/L$ \\ error \%\end{tabular} \\ \specialrule{.1em}{.05em}{.05em}
			A    & 6.66    & 0.541   & 0.542 & 0.23                                                               \\ 
			B    & 7.37    & 0.538   & 0.524 & -2.68                                                               \\ 
			C    & 8.10    & 0.523   & 0.528 & 0.96                                                               \\ 
			D    & 8.17    & 0.534   & 0.541 & 1.27                                                               \\ 
			E    & 7.54    & 0.561   & 0.562 & 0.27                                                               \\ 
			F    & 7.78    & 0.568   & 0.572 & 0.78                                                               \\ 
			\specialrule{.1em}{.05em}{.05em}
		\end{tabular}
	\end{table}
	
	\noindent In addition, the reliability of CFD-PBM was also assessed by the predicted pressure gradient comparison with field data, as shown in Table.~\ref{tab:validation}. It is evident from Table.~\ref{tab:validation} that the predicted pressure gradient is in excellent agreement with field data with a maxim error lower than 3\%. It was worth noting that the CFD-PBM model accuracy significantly improved in prediction the pressure compared to our earlier results~\citep{sontti2022computational}. The improvement in the CFD-PBM model prediction in pressure drop and velocity distribution with field data may be due to taking into account bitumen droplet breakage and aggregation phenomena. A CFD model with constant bitumen droplet sizes and a CFD-PBM model prediction are compared with experimental pressure drop and velocity data. With experimental field data, CFD-PBM results showed minimum deviations compared to the CFD model predictions as shown in Table. S2. \\
	
	The sensitivity of the number of bins is systematically studied while keeping the minimum and maximum size of the bitumen droplet range as shown in Table \ref{tab:Bitumeninitialsize} for all cases. There is a negligible change in pressure drop for four bin configurations (Table. S3). Moreover, the predicted pressure drop is also compared to the field data, and the error percentage remains consistent. However, the velocity profile along the lines exhibits a similar trend (Fig.S2), but the average velocity error percentage is reduced when comparing to the field data. This results in significantly increased computational time. Taking this into consideration, we attempt to use six bins.

	\section{Results and discussion}
	
	\subsection{Effect of slurry velocity on the distribution of bitumen droplets}
	In this section, we systematically investigate the effect of slurry velocity on the distribution of solid particles and bitumen droplets, for a range of velocities from 3.62 to 6.62 m/s. We present results including the evolution of bitumen droplet diameter and profiles of bitumen droplets within the pipeline.\\
	
	
	\begin{figure}[!ht]
		\centering
		\includegraphics[width=0.5\textwidth]{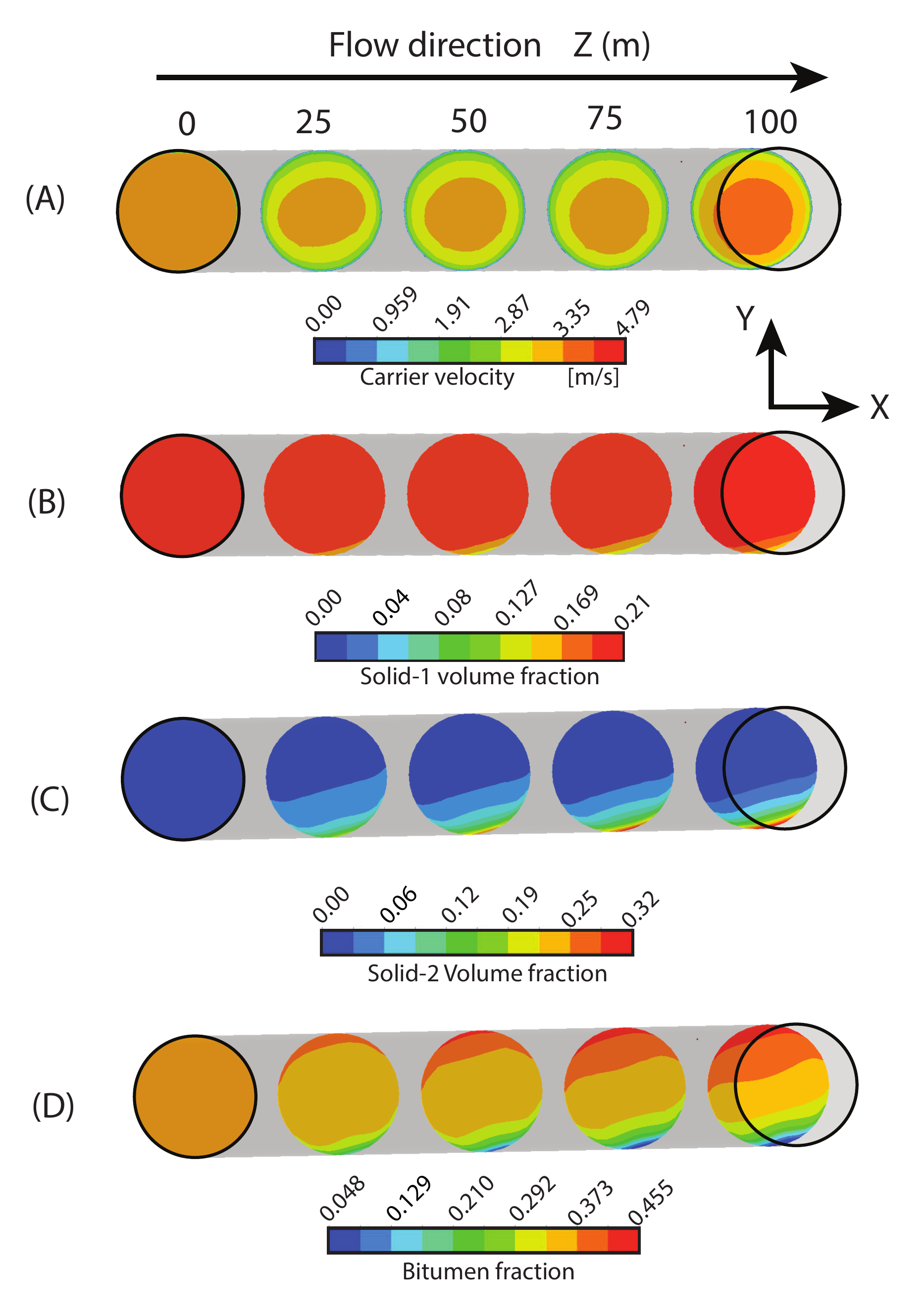}
		\caption{(A) Carrier velocity, (B) solid-1 fraction, (C) solid-2 fraction, and (D) bitumen distribution along the Z-direction for the mixture velocity of $V_{m}=3.62 \, \text{m/s}$ at flow time $t=200 \, \text{s}$.}
		\label{fig:Velocity1}
	\end{figure}
	
	Figure \ref{fig:Velocity1} shows cross-sectional views of carrier velocity, solid fractions, and bitumen fractions at five positions along the length of the pipe for a slurry velocity of 3.62 m/s. As seen in Fig. \ref{fig:Velocity1}A, the maximum carrier velocity is observed in the middle of the pipe, gradually increasing in magnitude from the inlet to the outlet. The distribution of solid concentration is also analyzed, revealing that small particles (75 $\mu$m) are uniformly distributed across the cross-section of the pipe from top to bottom, as shown in Fig. \ref{fig:Velocity1}B. This can be attributed to the relatively small effect of gravitational force on small particles. However, there is a slight change in the distribution of small particles along the length of the pipe, particularly in the region from Z=50 to 100 m, as shown in Fig. \ref{fig:Velocity1}B. , where coarse particles tend to accumulate at the bottom of the pipe, owing to the gravitational force acting on them.\\
	
	Furthermore, Fig. \ref{fig:Velocity1}C indicates the accumulation of large particles at the bottom of the pipe due to their high mass and the strong gravitational force acting on them. At the bottom of the pipeline, the sand particles follow the fluid velocity, but with a slightly lower velocity~\citep{li2018hydrodynamic}. The asymmetric distribution of solids in horizontal pipelines resulted from the particle settling and interactions between particles in the lower portion.  The sand particle size, particle density, and turbulence characteristics also play a significant role in settling and asymmetric distribution. Importantly, the non-Newtonian behavior and gravity-induced settling contribute to uneven particle concentration as a result asymmetric distribution observed in the perpendicular direction \citep{zheng2021turbulent}. The asymmetric disposition has been observed in the oil-sands industry~\citep{shook2013slurry}.
	
	Finally, Fig. \ref{fig:Velocity1}D shows the distribution of bitumen droplets along the length of the pipe, revealing that initially, they are mainly distributed from top to bottom of the pipe at Z=50 m. However, as the droplet size increases to the average size, the droplets start accumulating from the top to the middle of the pipe. Overall, the results indicate that the distribution of particles and bitumen droplets varies along the length of the pipe due to the interplay of gravitational force, carrier velocity, and particle sizes also wall interactions.\\
	
\begin{figure}[!ht]
		\centering
		\includegraphics[width=0.8\textwidth]{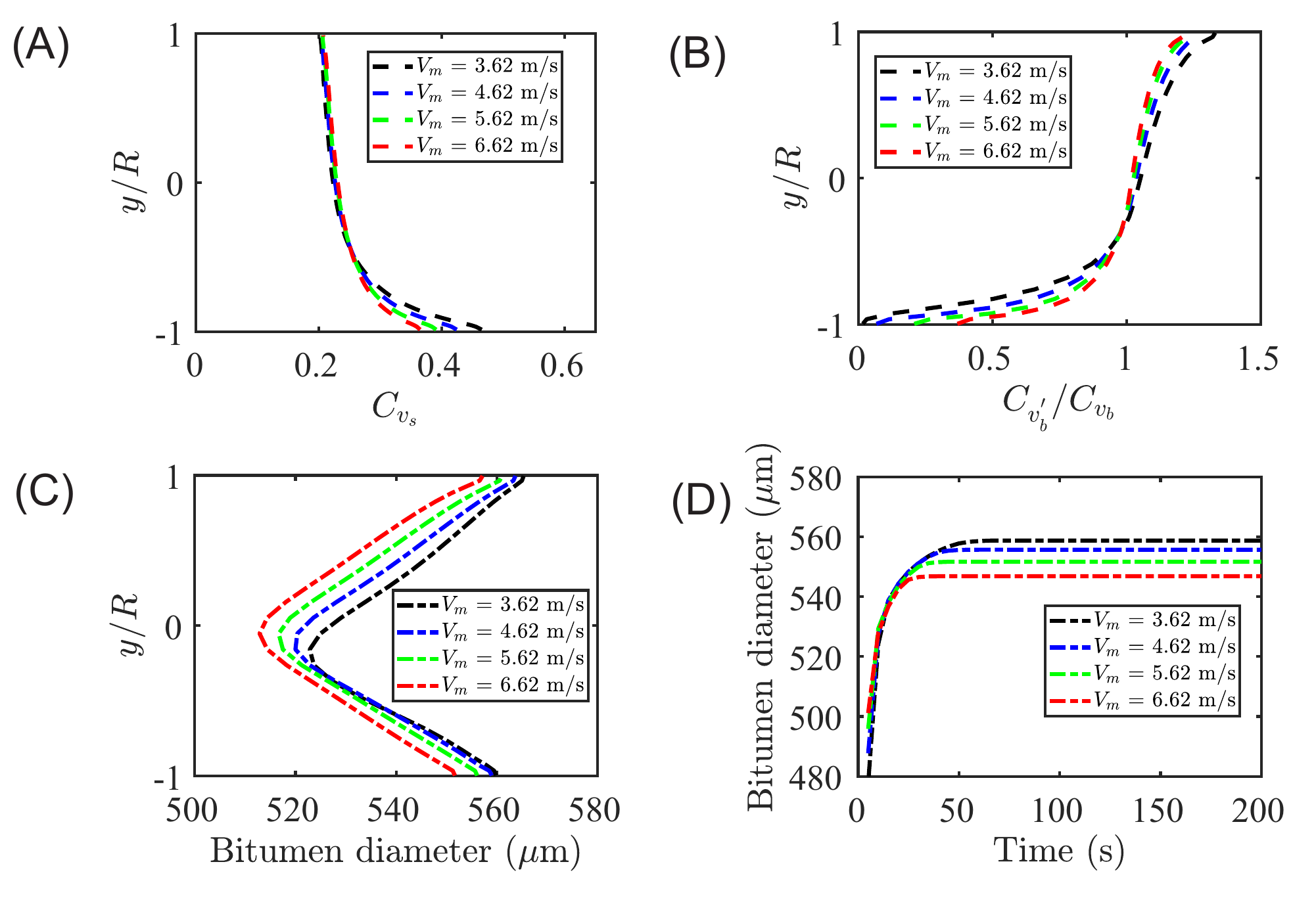}
		\caption{\label{fig:Velocity2} Effect of mixture velocity on (A) chord-average total solid concentration profiles on plane YX at Z=100 m, (B) bitumen distribution profiles along the vertical reference line at Z-100 m, (C) bitumen droplet diameter profile along the vertical reference line at Z-100 m, and (D) bitumen droplet diameter evolution with flow time on a reference point at Z= 100 m and Y= 0.30 m.}
	\end{figure}

	The influence of slurry velocity on the chord-average total solid concentration and bitumen distribution profiles was analyzed at Z=100 m for a range of velocities, as shown in Fig.\ref{fig:Velocity2}A and B. It is evident from the solid concentration profiles, velocity has shown a minimal impact on the distribution of solids, especially from the top to the middle. However, at the bottom of the pipe, slurry velocity significantly influenced the solid concentration distribution. For lower slurry velocities, a higher solid concentration was found. Despite this, for similar conditions, with an increase in the slurry velocity, significant differences in concentration distribution were observed, and for higher velocities, the solid concentration was relatively lower compared to smaller slurry velocities (Fig.\ref{fig:Velocity2}A). This is mainly due to the dispersion of coarse particles at higher slurry velocity. Therefore, with an increase in slurry velocity, the solid concentration distribution changes.\\
	
	Figure \ref{fig:Velocity2}B shows the bitumen distribution profiles along the vertical reference line at Z=100 m for different slurry velocities. The CFD-PBM results indicate that the slurry velocity range also has a significant influence on the distribution of bitumen concentration. At lower slurry velocities, bitumen droplets are distributed from the top to bottom part of the pipe. However, at higher slurry velocities, the bitumen concentration distribution significantly changed due to an increase in turbulence in the pipeline.\\
	
	The bitumen droplet diameter profiles are analyzed for different slurry velocities at a special location (ie., Z=100 m) along the vertical reference line and at a fixed probe Z=100 m and Y=0.30 m from the center of the pipe as shown in Fig. \ref{fig:Velocity2}C and D. From the CFD-PBM results, it is indicated that the bitumen aggregated droplet size has shown a notable difference at the center of the pipeline for the considered slurry velocity conditions. For higher slurry velocity conditions, the aggregated droplet size is relatively small at the center and the profile also significantly shifted backward as shown in \ref{fig:Velocity2}C. The reason for this might be that breakage frequency efficiency played a dominant role at higher velocities, due to the increase in particle-bitumen interactions. As a result, the aggregated bitumen droplet size is decreased at the center. However, for lower slurry velocity the bitumen droplet size showed a maximum value where the breakage frequency might have a minimal impact on bitumen droplets. The bitumen droplet distribution has shown symmetrical profiles in the vertical direction. It suggests that the droplets are almost evenly distributed in the pipe from the top to bottom.\\

	\begin{figure}[!ht]
		\centering
		\includegraphics[width=0.8\textwidth]{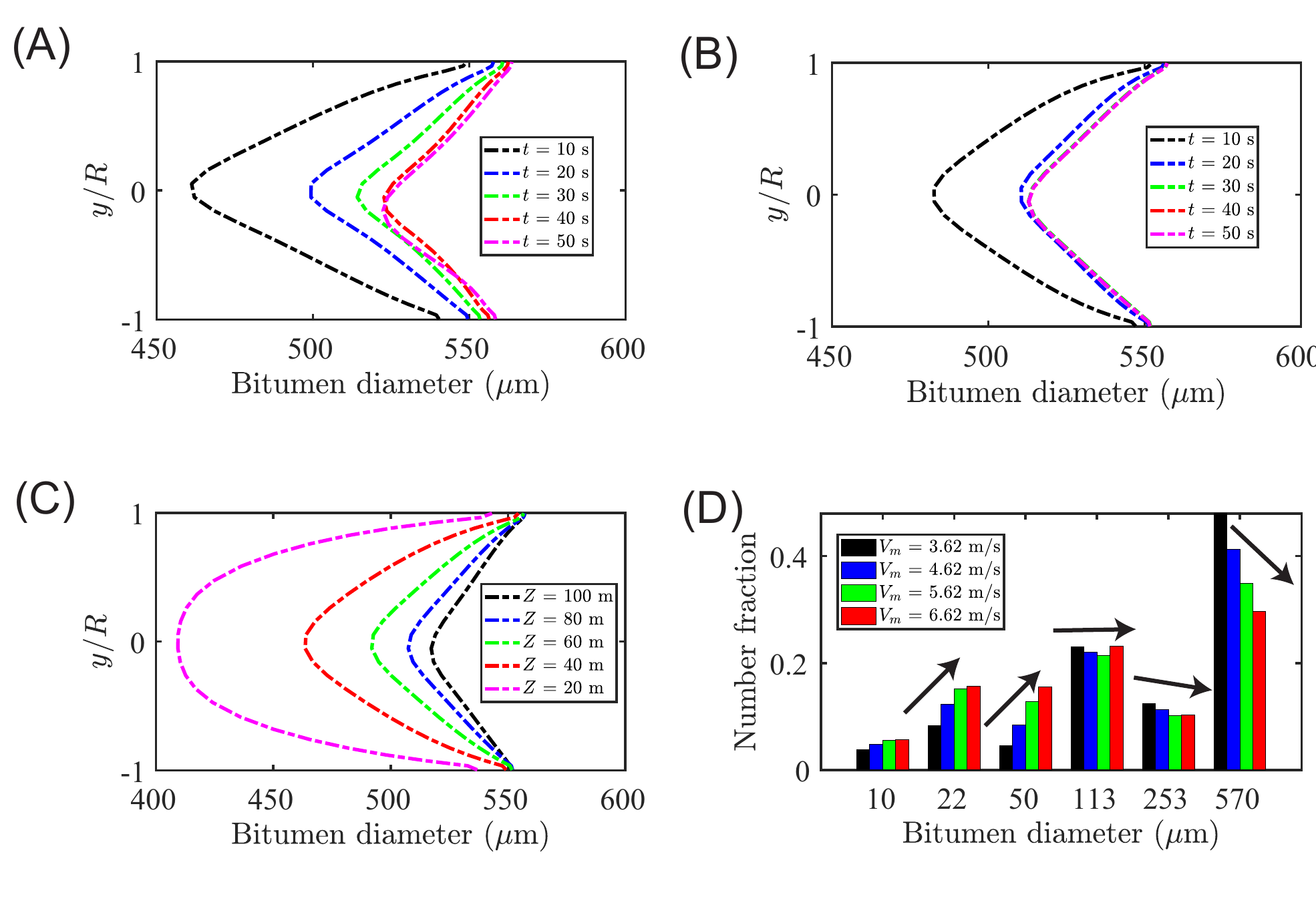}
		\caption{\label{fig:Velocity3} Bitumen droplet diameter profiles for different flow times at Z=100 m along the $YY^{'}$ vertical line (A) for the mixture velocity of $V_{m}$= 3.62 m/s, (B) for the mixture velocity of $V_{m}$= 6.62 m/s. (C) Bitumen droplet diameter profiles at the different lengths for the mixture velocity of $V_{m}$= 6.62 m/s at a fixed flow time t= 200 s. (D) Effect of mixture velocity on initial bitumen drop size distribution at Z=100 m at a fixed flow time t= 200 s. }
	\end{figure}
	
	At a fixed probe point with coordinates Z=100 m and Y=0.30 m, the evolution of bitumen droplet diameter with the flow time is depicted in Figure \ref{fig:Velocity2}D. The results demonstrate that as slurry velocity increases, the size of aggregated bitumen droplets at the top section of the pipe decreases from intense turbulence dispersion. At higher velocities, there is a greater probability for droplets to collide and coalesce, which can lead to the formation of larger droplets. It is also possible that higher velocities can cause larger droplets to break up into smaller ones. It is evident that higher velocities cause the aggregated bitumen droplet size to decrease more rapidly, reaching a diameter of t=30 s. The bitumen droplet diameter gradually increased until it reached t=50 s, after which it reached a stable size. As a result of this observation, the bitumen droplet diameter size reached a stable value at t=50 s.\\
	
	
	\noindent The evolution of bitumen droplet sizes at low and high velocities is particularly interesting. A vertical reference line is used to analyze the profiles of bitumen droplet diameter at different flow times, to describe the evolution of bitumen droplet diameter over time. Figure \ref{fig:Velocity3} A and B show the bitumen droplet diameter profiles for $V_m$ = 3.62 m/s and $V_m$ = 6.62 m/s, respectively. A significant difference is observed in the bitumen droplet size as the flow time changes, with the droplet diameter gradually increasing. The bitumen droplet diameter profiles obtained by CFD-PBM models indicate that the droplet profiles gradually reach stability at specific positions, based on the slurry velocity conditions. The smaller droplets tend to exhibit higher velocities within the slurry mixture. This behavior can be attributed to the breakup of larger droplets and the limited coalescence of smaller droplets during the transport process. The hydrodynamic forces acting on the droplets become more significant at higher flow velocities, inducing the breakup of larger droplets into smaller fragments. Additionally, the limited residence time at higher velocities hinders the coalescence of smaller droplets. As a result, the smaller droplets remain relatively intact and continue to travel at higher velocities.\\
	
	\begin{figure}
		\centering
		\includegraphics[width=0.8\textwidth]{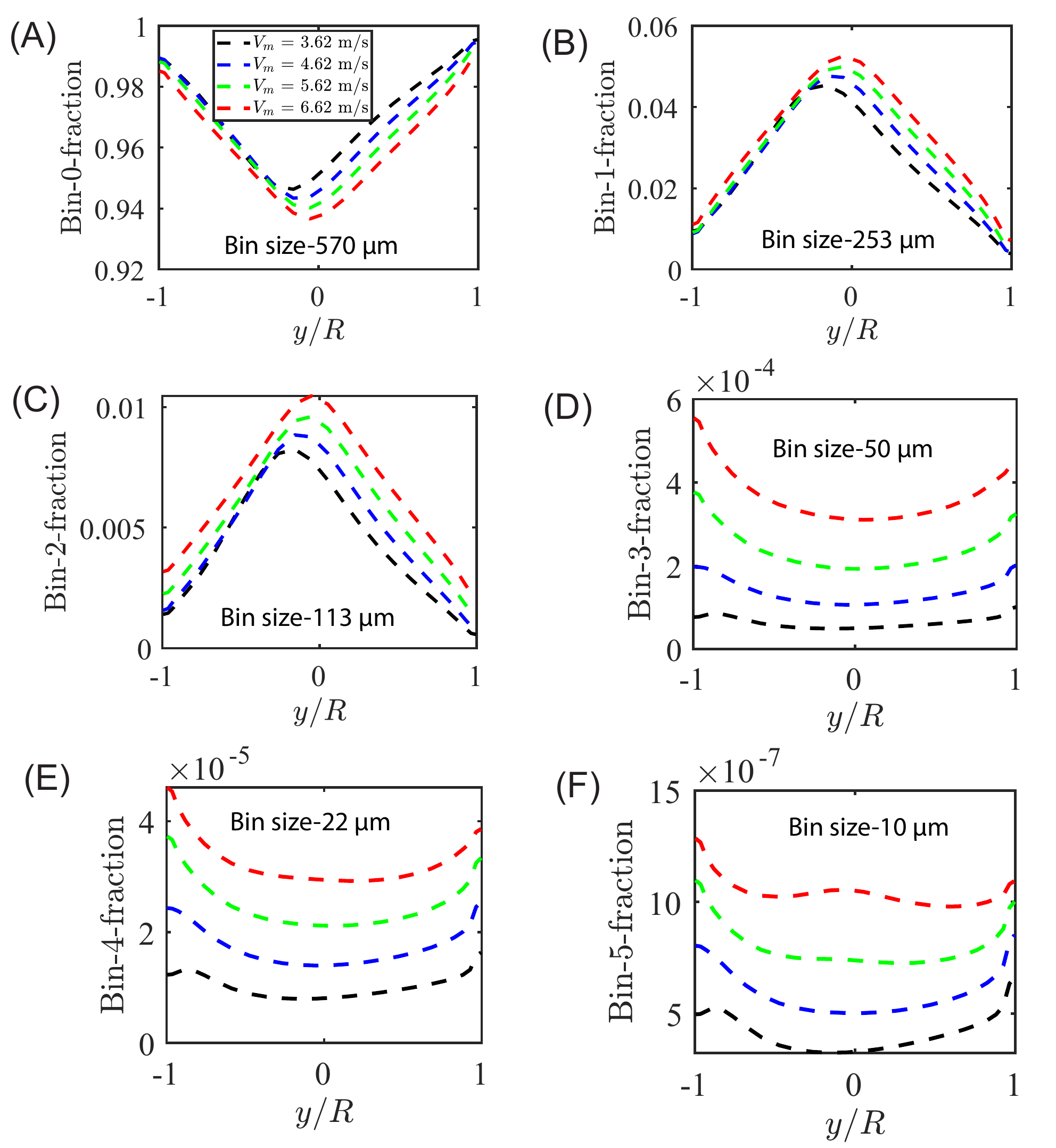}
		\caption{\label{fig:Velocity4} (A)\textendash (F) Effect of mixture velocity on the volume fraction of bins for different bins along the vertical reference line at Z=100 m and flow time t= 200 s. }
	\end{figure}
	
	To investigate the bitumen droplet diameter profiles at different locations, various vertical reference lines are considered to analyze the profiles from Z = 20 m to Z = 100 m. As shown in Fig. \ref{fig:Velocity3}C, the bitumen droplet diameter profile shifts from a lower diameter size to a higher one as the pipe length progresses. A symmetrical change in bitumen droplet size is observed  along the length of the pipe and near the outlet of the transport system, which indicates stable conditions near the outlet. The bitumen droplet size distribution is analyzed across the cross-section at Z= 100. The bitumen droplet distribution can be represented in terms of the number of fractions (i.e., the ratio of discrete number density to total number density) as shown in Fig.\ref{fig:Velocity3}D. In a highly turbulent flow near the wall, larger droplets in the boundary layer may experience reduced shear forces. As a result, the number fraction of larger droplets can decrease, while the number fraction of smaller droplets can increase near the pipe inlet. However, if the smaller droplets tend to coalesce or aggregate, they may reform larger droplets near the pipe outlet, leading to an increase in the number fraction of larger droplets. \\  
	
	\noindent In addition, the volume fraction profiles of all the bitumen droplet size bins are analyzed for different slurry velocities, as shown in Figure \ref{fig:Velocity4}. The results indicate that the slurry velocity has a significant effect on the bitumen droplet distribution. For larger-size bins, the maximum bin function is found at the top and bottom walls of the pipe (Figure \ref{fig:Velocity4}A). The high-velocity flow at the center of the pipe creates high shear forces and turbulence that break up larger droplets into smaller ones, causing a decrease in the number fraction of larger droplets and an increase in the number fraction of smaller ones near the center of the pipe (Figure \ref{fig:Velocity4}B and C). The maximum volume fractions of bin-253 $\mu$m and bin-113 $\mu$m are observed at the center of the pipe, where the flow field develops the highest velocity, indicating that droplets in these bins are preferentially located in this region. In contrast, for smaller size bins, the bin fraction is more uniform across the pipeline with an slight higher fraction near the walls of the pipe (Figure \ref{fig:Velocity4}D-F). This is likely due to the interactions between the small bitumen droplets and solid particles leading to uniform dispersion. 
	
	
	\subsection{Effect of bitumen fraction on the distribution of bitumen droplets}
	
\begin{figure} [!ht]
		\centering
		\includegraphics[width=0.5\textwidth]{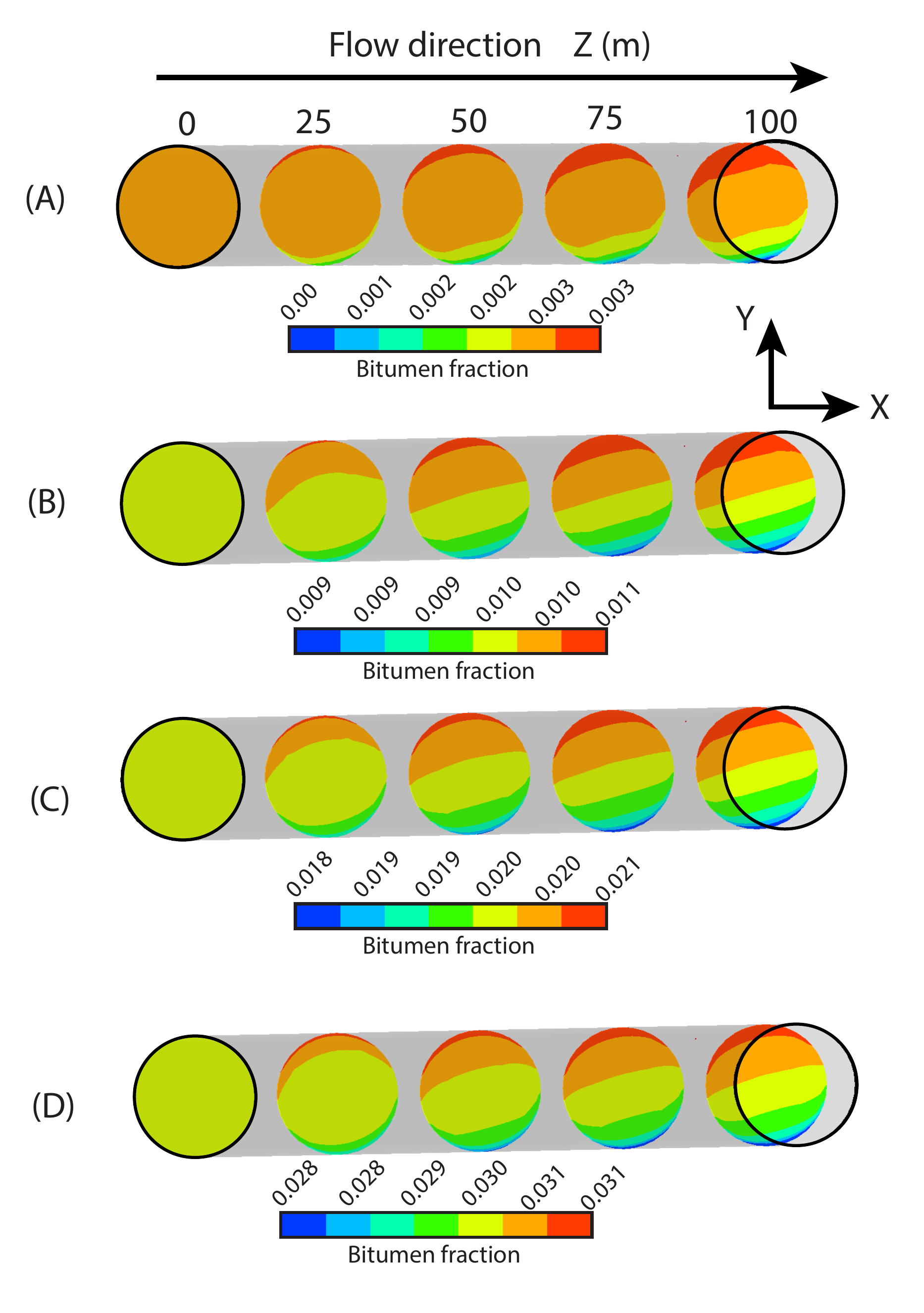}
		\caption{\label{fig:Bitumenfrac1} Bitumen distribution along the Z-direction (A) bitumen fraction-0.0025, (B) bitumen fraction-0.01, (C) bitumen fraction-0.02, and (D) bitumen fraction-0.03 for a mixture velocity of $V_{m}$= 5.62 m/s at flow time t = 200 s.}
	\end{figure}

	In order to examine the impact of bitumen fraction on tailings slurry systems, different bitumen compositions between 0.0025 and 0.03 are studied systematically, which represents a typical range of bitumen fraction values found in industrial tailings residuals. Fig. \ref{fig:Bitumenfrac1} shows contour plots of bitumen droplet distribution along the flow direction for different bitumen fractions. CFD-PBM numerical results indicate that for lower bitumen fraction conditions, bitumen droplets are distributed evenly from the top part of the pipe to the center of the pipe (Fig. \ref{fig:Bitumenfrac1}A). The results also reveal that the turbulent flow characteristics of the slurry system are affected by the bitumen fraction. As shown in Fig. \ref{fig:Bitumenfrac1}B-D, the accumulation of bitumen at the top part of the pipe with an increase in bitumen fraction can be explained by the difference in the densities between the bitumen droplets and the solid particles. Additionally, the turbulence and flow patterns in the pipe can also influence the distribution of bitumen droplets.\\
	
\begin{figure} [!ht]
		\centering
		\includegraphics[width=0.8\textwidth]{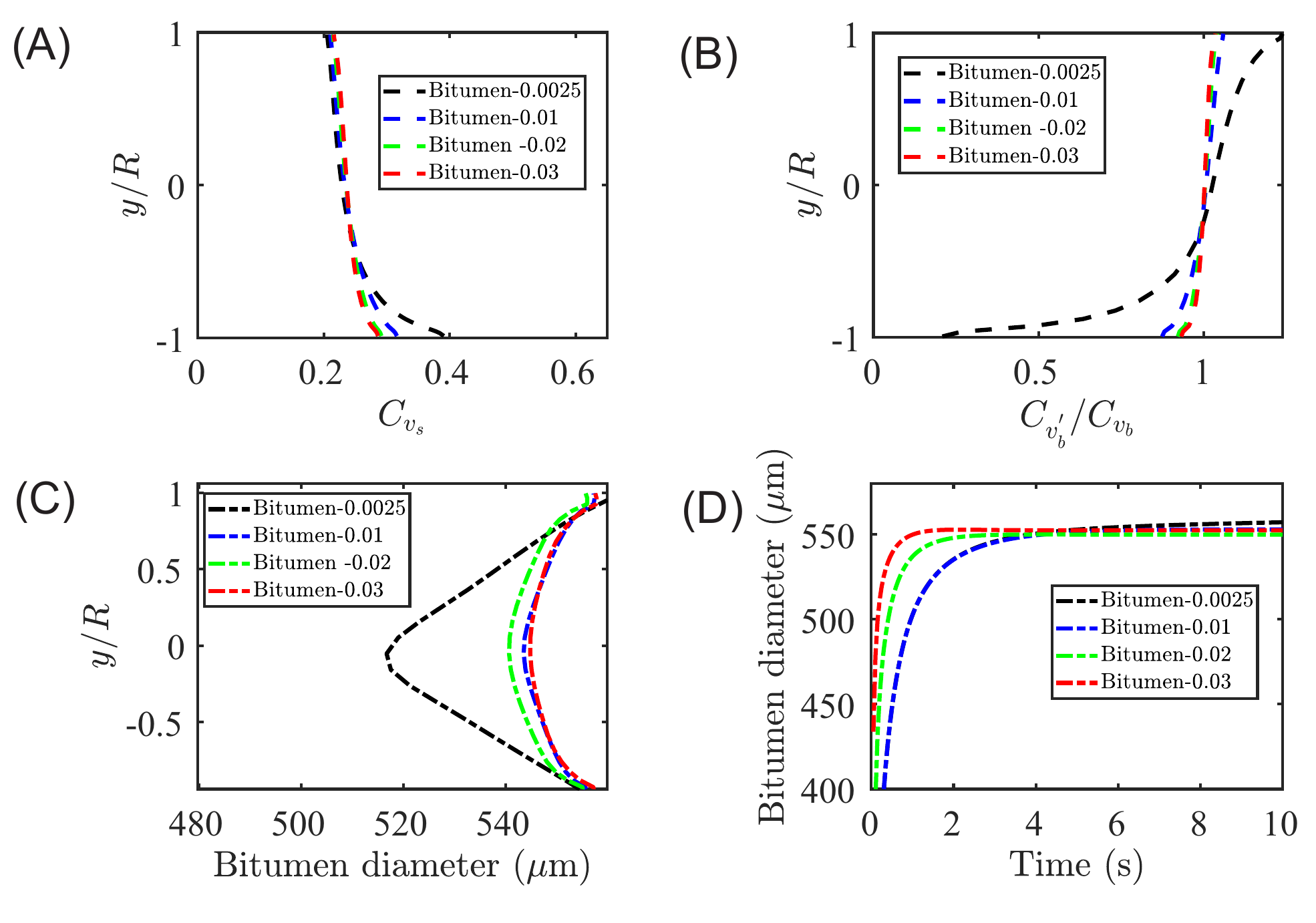}
		\caption{\label{fig:Bitumenfrac2} Effect of bitumen fraction on (A) chord-average total solid concentration profiles on plane YX at Z=100 m, (B) bitumen distribution profiles along the vertical reference line at Z-100 m, and (C) bitumen droplet diameter profile along the vertical reference line at Z-100 m at flow time t = 200 s. (D) Bitumen droplet diameter evolution with flow time on a reference point at Z= 100 m and Y= 0.30 m for a mixture velocity of $V_{m}$= 5.62 m/s.}
	\end{figure}

	\noindent As demonstrated in Fig.\ref{fig:Bitumenfrac2}A, With an increase in bitumen fraction from 0.0025 to 0.03, the solids concentration profile distribution in a circular pipeline shows a noticeable variation in the vertical direction. The CFD-PBM simulation results reveal that an increase in bitumen fraction causes the distribution of coarse particles to shift upwards from the bottom to the top part of the pipe, while the distribution of small particles remains nearly constant across the pipe. The concentration of fine particles gradually increases in the upper half of the pipe, and the asymmetry of the fine particle concentration profile along the vertical direction increases. This behavior can be attributed to the impact of bitumen droplets on the fluid flow and the interaction forces between bitumen droplets and solid particles.\\
	
	The bitumen concentration profile distribution along the vertical direction changes with an increase in bitumen fraction as shown in Fig.\ref{fig:Bitumenfrac2}B. At lower bitumen fractions (i.e., 0.0025), the bitumen droplets are distributed evenly from the top to the center of the pipe, with a slightly higher concentration towards the top. As the bitumen fraction increases, the concentration of bitumen droplets increases throughout the pipe and a larger proportion of the droplets are found at the top of the pipe.  This is due to the fact that bitumen droplets are lighter than solid particles and can be transported more easily by the flow. Additionally, the bitumen droplets tend to aggregate and coalesce due to their hydrophobic nature, leading to larger droplets at the top of the pipe.\\
	
	\noindent As shown in Fig.\ref{fig:Bitumenfrac2}C, with an increase in bitumen concentration from 0.0025 to 0.03, the CFD-PBM simulation results show a clear trend of decreasing bitumen droplet diameter along the vertical direction of the circular pipe from the top to center of the pipe and then gradually increasing center to the bottom of the pipe. At the lowest bitumen concentration of 0.0025, the symmetry in the bitumen droplet diameter profile along the vertical direction becomes more pronounced as the bitumen concentration decreases, with a sharper peak near the center of the pipe. However, as the bitumen concentration increases to 0.03, the bitumen droplet diameter increases along the vertical direction, particularly near the top of the pipe. This trend can be attributed to the increased coalescence and breakup of larger bitumen droplets due to the higher local shear rates and turbulence in the upper regions of the pipe, resulting in a higher concentration of smaller droplets in this region.\\
	
	At the fixed point near the top wall, the bitumen droplet diameter profile varies with flow time as shown in Fig.\ref{fig:Bitumenfrac2}D. Initially, at t=0 s, the droplet diameter distribution is nearly uniform throughout the pipe. As the flow progresses, the droplet diameter profile shows a significant change due to the agglomeration of small droplets into larger droplets. This agglomeration process is due to the collision of droplets under the influence of turbulence in the slurry flow. For higher bitumen concentration, after t=2 s, the droplet diameter profile reaches a steady state faster than the lower bitumen concentration. \\
	
	\begin{figure} [!ht]
		\centering
		\includegraphics[width=0.8\textwidth]{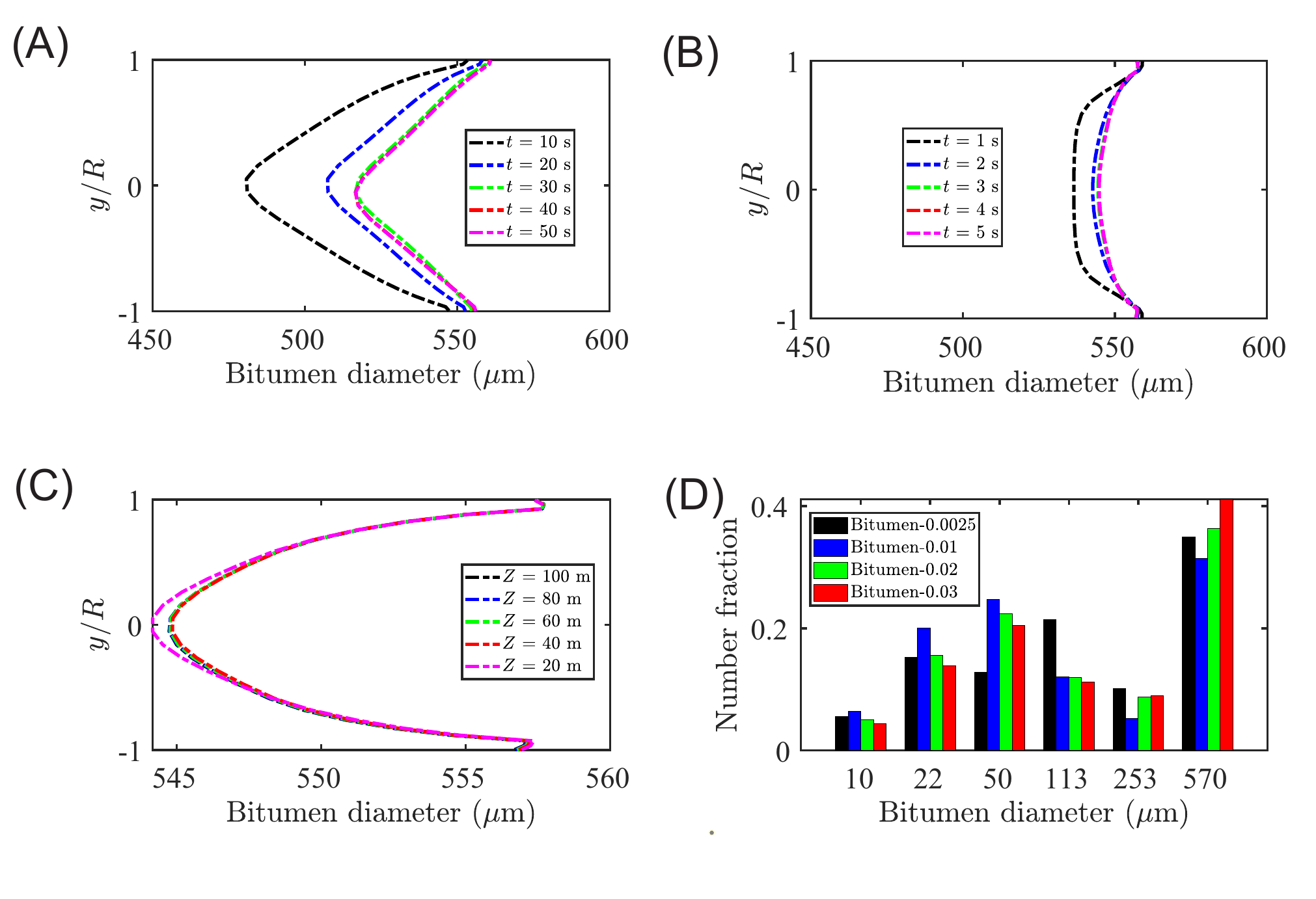}
		\caption{\label{fig:Bitumenfrac3}  Bitumen droplet diameter profiles for different flow times at Z=100 m along the $YY^{'}$ vertical line (A) bitumen fraction-0.0025, (B) bitumen fraction-0.03, (C) bitumen droplet diameter profiles at the different lengths for bitumen fraction 0.03 case at a fixed flow time t= 200 s. (D) Effect of bitumen fraction on initial bitumen drop size distribution at Z=100 m at a fixed flow time t= 200 s and a mixture velocity of $V_{m}$= 5.62 m/s.}
	\end{figure}
	
	
	The bitumen droplet diameter profiles along a vertical line are analyzed for two extreme bitumen concentrations: 0.0025 and 0.03, and at flow times ranging from 10s to 50s. For the lower bitumen concentration, at 10s, the droplet size distribution is uniform throughout the vertical line and the aggregate bitumen droplet size also smaller at the center of the pipe as displayed in Fig.\ref{fig:Bitumenfrac3}A. As time progresses significant change in the bitumen diameter profile is observed and the droplets of larger sizes accumulate near the top and bottom of the pipe, while smaller droplets remain uniformly distributed in the pipeline. In contrast, a noticeable difference is observed for t=1 s to t=2 s at higher bitumen concentrations as portrayed in Fig.\ref{fig:Bitumenfrac3}B. The results suggest that the bitumen concentration has a significant impact on the droplet size distribution along the vertical line and its evolution with time.  In addition, along the length of the pipe, the bitumen diameter profiles are also analyzed, and the CFD-PBM results demonstrated that after reaching the Z=40m the negligible change in bitumen diameter profiles along the vertical direction as can be seen in Fig.\ref{fig:Bitumenfrac3}C. The number fraction analysis demonstrated that  with an increase in bitumen concentration larger aggregates are observed as depicted in Fig.\ref{fig:Bitumenfrac3}D. consistent with the distribution in Fig.\ref{fig:Velocity3}D. \\    
	
	
	\subsection{Effect of bubble size on the distribution of bitumen droplets}
\begin{figure} [!ht]
		\centering
		\includegraphics[width=0.5\textwidth]{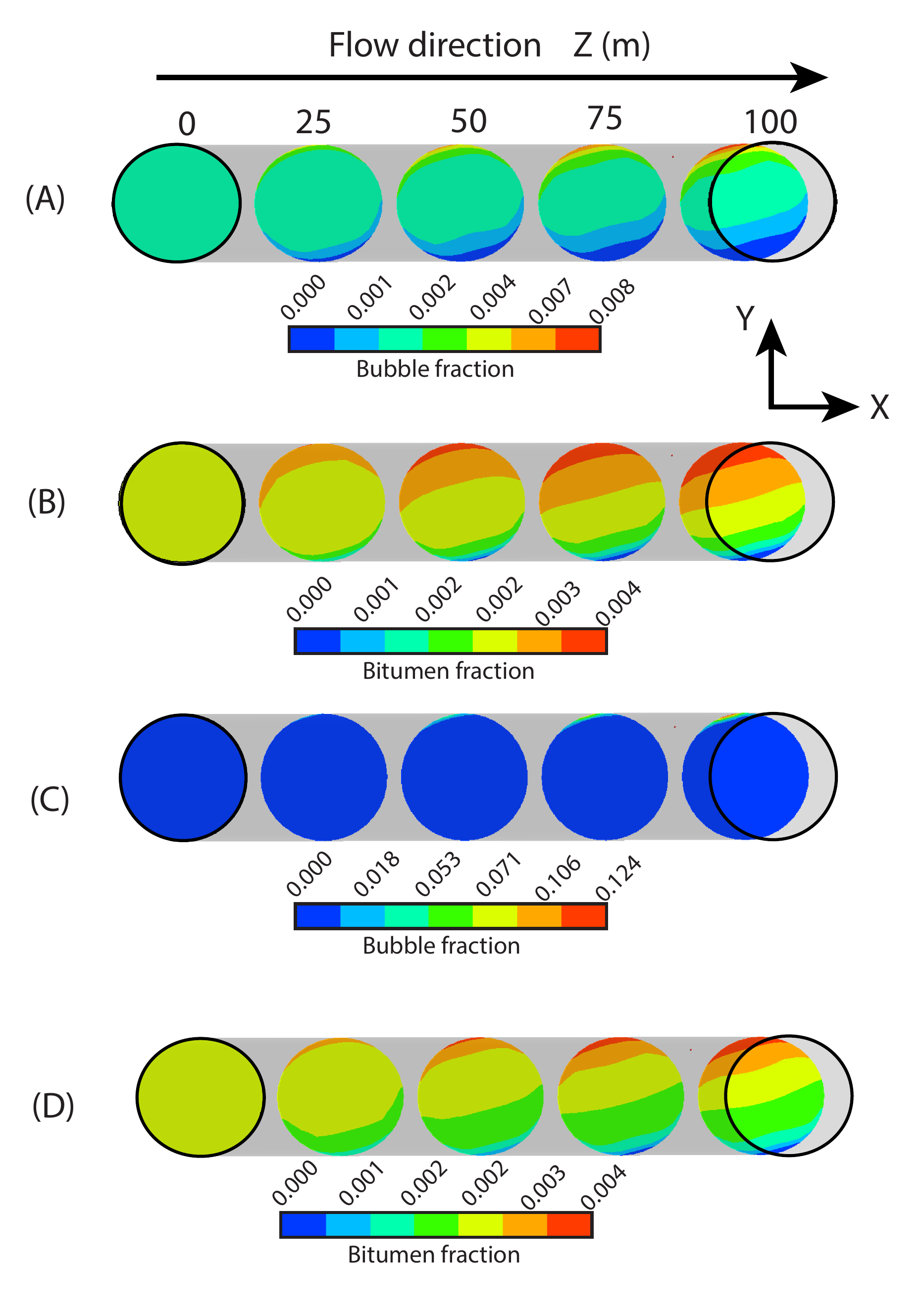}
		\caption{\label{fig:Bublbesize1} (A) Bubble fraction distribution along the Z-direction and (B) bitumen fraction for bubble size 250 $\mu$m. (C) Bubble fraction distribution along the Z-direction and (D) bitumen fraction  for bubble size 1000 $\mu$m and a mixture velocity of $V_{m}$= 5.62 m/s at flow time t = 200 s.}
	\end{figure}

	In this section, gas bubbles with sizes of 250, 500, 750, and 1000 $\mu$m are added to the slurry system based on the literature data of bubble size rage \cite{zahid2020experimental, rosas2018measurements}. The gas bubble fraction is considered the same as the bitumen fraction for all the cases. The distribution of both bitumen and bubbles along the length of the pipe at different locations is analyzed for bubble sizes of 250 and 1000 microns as shown in Fig.\ref{fig:Bublbesize1}A-D. The results show that, for the smaller bubbles (250 $\mu$m), the distribution is relatively uniform along the length of the pipe, with some accumulation at the upper part (Fig.\ref{fig:Bublbesize1}A). However, for the larger bubbles (1000 $\mu$m), the distribution becomes more heterogeneous, with a higher concentration at the top of the pipe due to buoyancy (Fig.\ref{fig:Bublbesize1}C).
	
	As for the bitumen, it is found that the distribution is influenced by the presence of bubbles. However, it is evident from Fig.\ref{fig:Bublbesize1}B, for smaller bubbles, the bitumen accumulates more at the top of the pipe, due to the upward flow of the gas phase. In comparison to larger bubbles, tiny bubbles attach more easily and faster to particles/droplets because of a smaller contact area. In the case of larger bubbles, the bitumen is mostly distributed at the top as depicted in Fig.\ref{fig:Bublbesize1}D. The distribution of bubbles and bitumen is affected by a combination of gravity and buoyancy forces, as well as the flow rate and size of the particles. The gravitational force exerted by small bubbles is usually negligible when compared to large bubbles\cite{parvathaneni2020role}. As a result, the buoyance force is more prominent in large bubbles to overcome the turbulence force.\\
	

	\begin{figure} [!ht]
		\centering
		\includegraphics[width=0.7\textwidth]{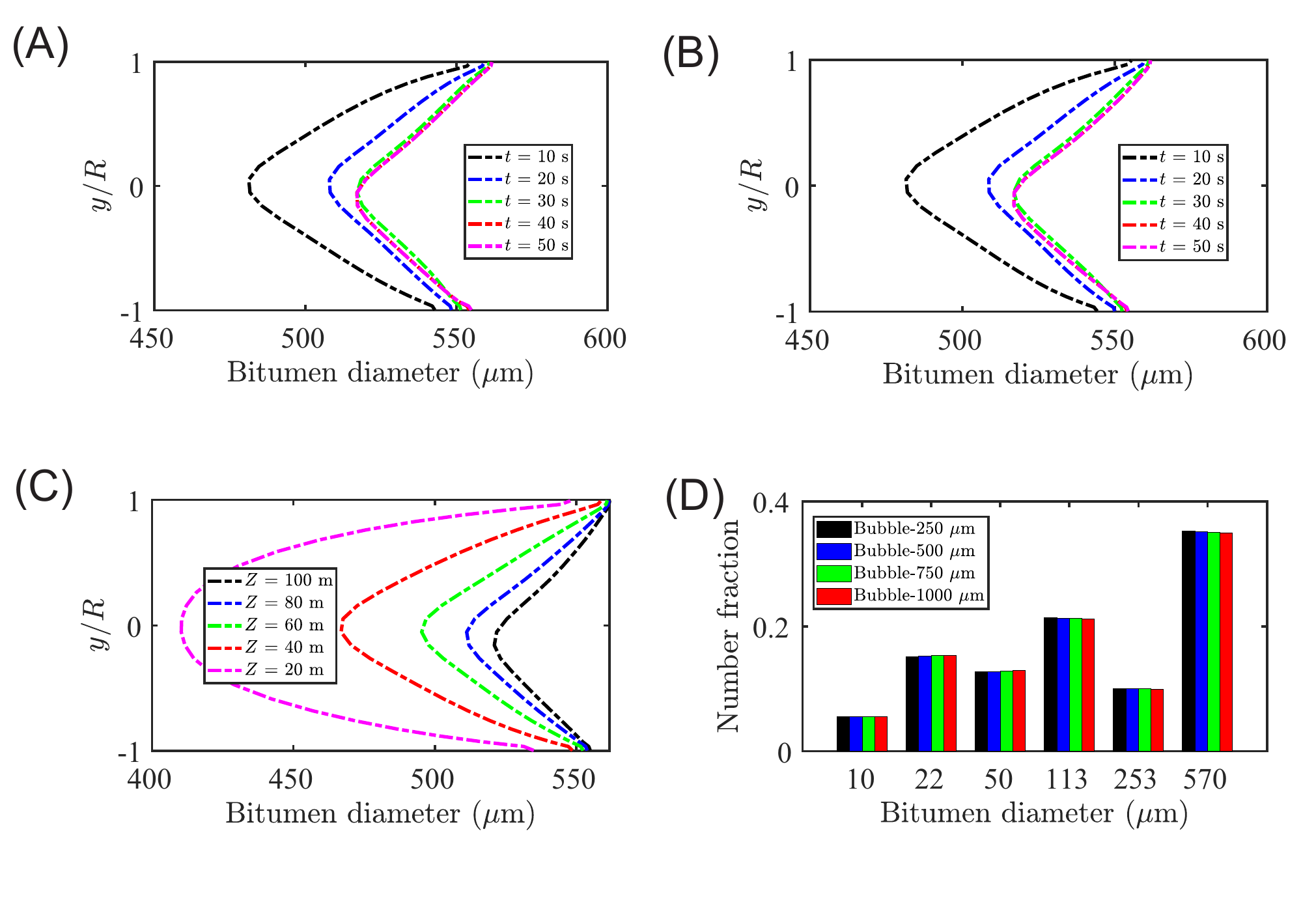}
		\caption{\label{fig:Bublbesize3}  Bitumen droplet diameter profiles for different flow times at Z=100 m along the $YY^{'}$ vertical line (A) bubble size- 250 $\mu$m, (B) bubble size- 1000 $\mu$m, (C) bitumen droplet diameter profiles at the different lengths for bubble size- 1000 $\mu$m case at a fixed flow time t= 200 s. (D) Effect of bubble size on initial bitumen drop size distribution at Z=100 m at a fixed flow time t= 200 s and a mixture velocity of $V_{m}$= 5.62 m/s.}
	\end{figure}
	
	The bitumen droplet diameter along the vertical reference line for two different cases are demonstrated in Fig.\ref{fig:Bublbesize3}A and B. The results of both cases showed similar trends and outcomes. This might be due to the consideration of the bubble fraction is relatively low for both cases. However, significant changes in the droplet diameter profile with length, and the droplet diameter increased due to aggregation near the outlet as shown in Fig.\ref{fig:Bublbesize3}C. The analysis showed that the bitumen droplet diameter increased as the length of the pipeline increased. This behavior can be attributed to the coalescence of the droplets, which leads to the formation of larger droplets. Furthermore, number fractions are also compared for all the cases and results indicated that bubble size has a negligible effect on the number fraction distribution (Fig.\ref{fig:Bublbesize3}D). The observed changes in the droplet diameter profile and the increase in droplet size due to aggregation could have important implications for pipeline transport operations, as they can affect the flow behavior and the efficiency of the pipeline system.
	
	\subsection{Effect of particle size on the distribution of bitumen droplets}
	The distribution of solid and bitumen fractions is influenced by combinations of particle sizes. Four groups of particle size combinations are studied with a mixture velocity of $V_{m}$=5.62m/s. Group 1 and Group 2 consisted of S1 particles with sizes of 200 $\mu$m and S2 particles with sizes of 500 $\mu$m, 700 $\mu$m, respectively. Group 3  and Group 4 included S1 particles with a size of 75 $\mu$m and S2 particles with sizes of 500 $\mu$m, 700 $\mu$m.\\
	
	\begin{figure} [!ht]
		\centering
		\includegraphics[width=0.7\textwidth]{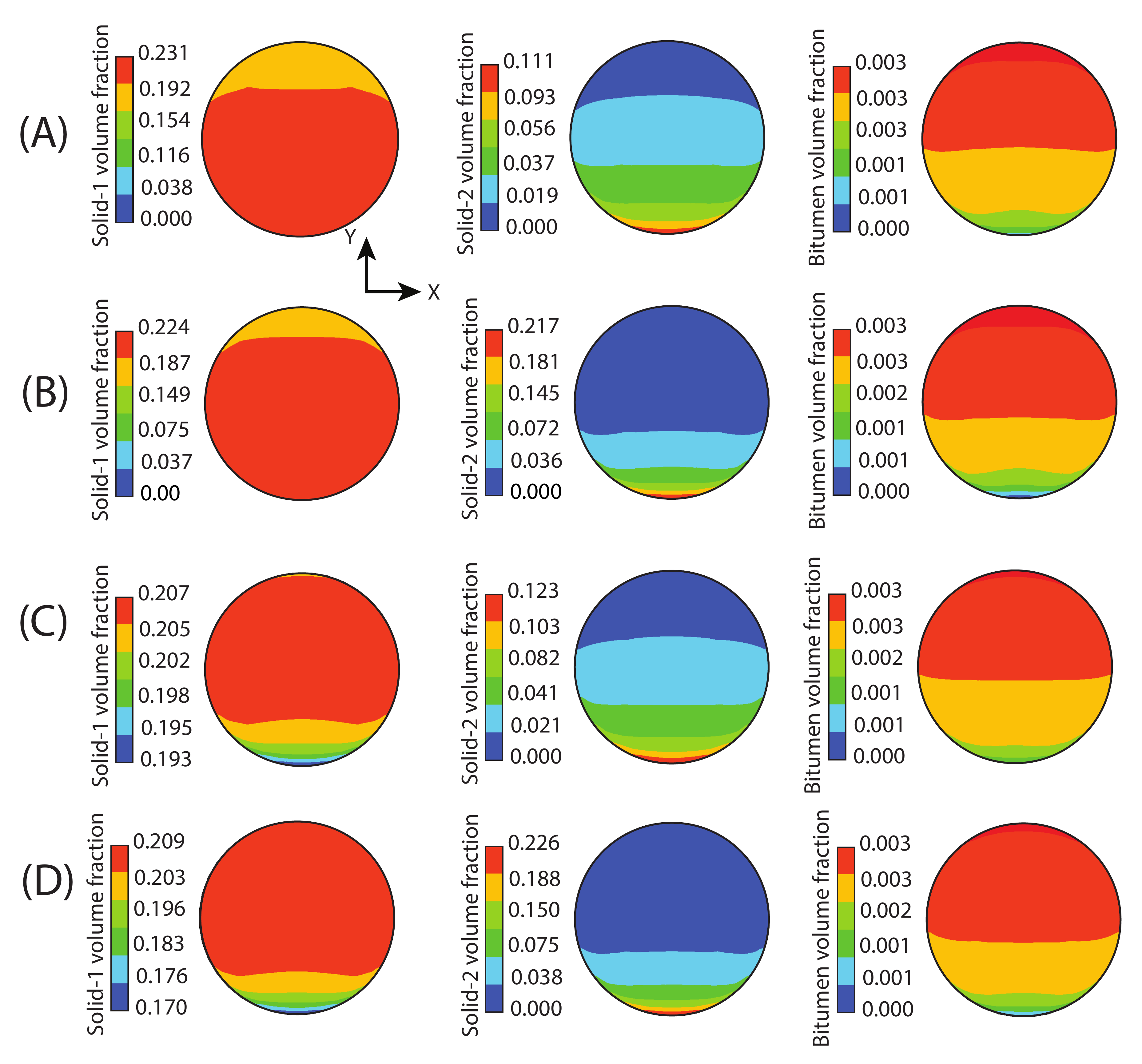}
		\caption{\label{fig:Partilce1} Effect of particle size combinations on solid and bitumen fraction distribution (A) group\textendash 1: S1\textendash 200 $\mu$m and S2\textendash 500 $\mu$m, (B) group\textendash 2: S1\textendash 200 $\mu$m and S2\textendash 700 $\mu$m, (C) group\textendash 3: S1\textendash 75 $\mu$m and S2\textendash 500 $\mu$m, and (D) group\textendash 4: S1\textendash 75 $\mu$m and S2\textendash 700 $\mu$m, at Z= 100 m along the vertical reference line for a mixture velocity of $V_{m}$= 5.62 m/s}
	\end{figure}
	
	Figure \ref{fig:Partilce1}A-D demonstrates the impact of particle size combinations on solid fraction and bitumen distribution at Z=100 m. CFD-PBM results reveal that, for group-1, smaller particles are distributed above the pipe center while larger particles are accumulated at the bottom of the pipe. Coarse particles are affected more by the gravitational force, leading to their accumulation at the bottom of the pipe. It is evident from Figure \ref{fig:Partilce1}A, the coarse fraction is minimal at the top and center of the pipe, whereas bitumen droplets and small particles are distributed throughout the pipe, and bitumen droplets are accumulated at the upper part of the pipe. A similar observation is seen for group-2 combinations, as shown in Fig.\ref{fig:Partilce1}B. However, group-3 and group-4 particle combinations showed significant differences in the solid concentration distribution where S1-75 $\mu$m particle (Fig.\ref{fig:Partilce1}C and D). The results indicate that small particles are distributed from the center to near the top wall of the pipeline. A noticeable change in coarse particle distribution is observed for group-4 particle combinations, as shown in Fig.\ref{fig:Partilce1}D. \\
	
	\begin{figure} [!ht]
		\centering
		\includegraphics[width=0.7\textwidth]{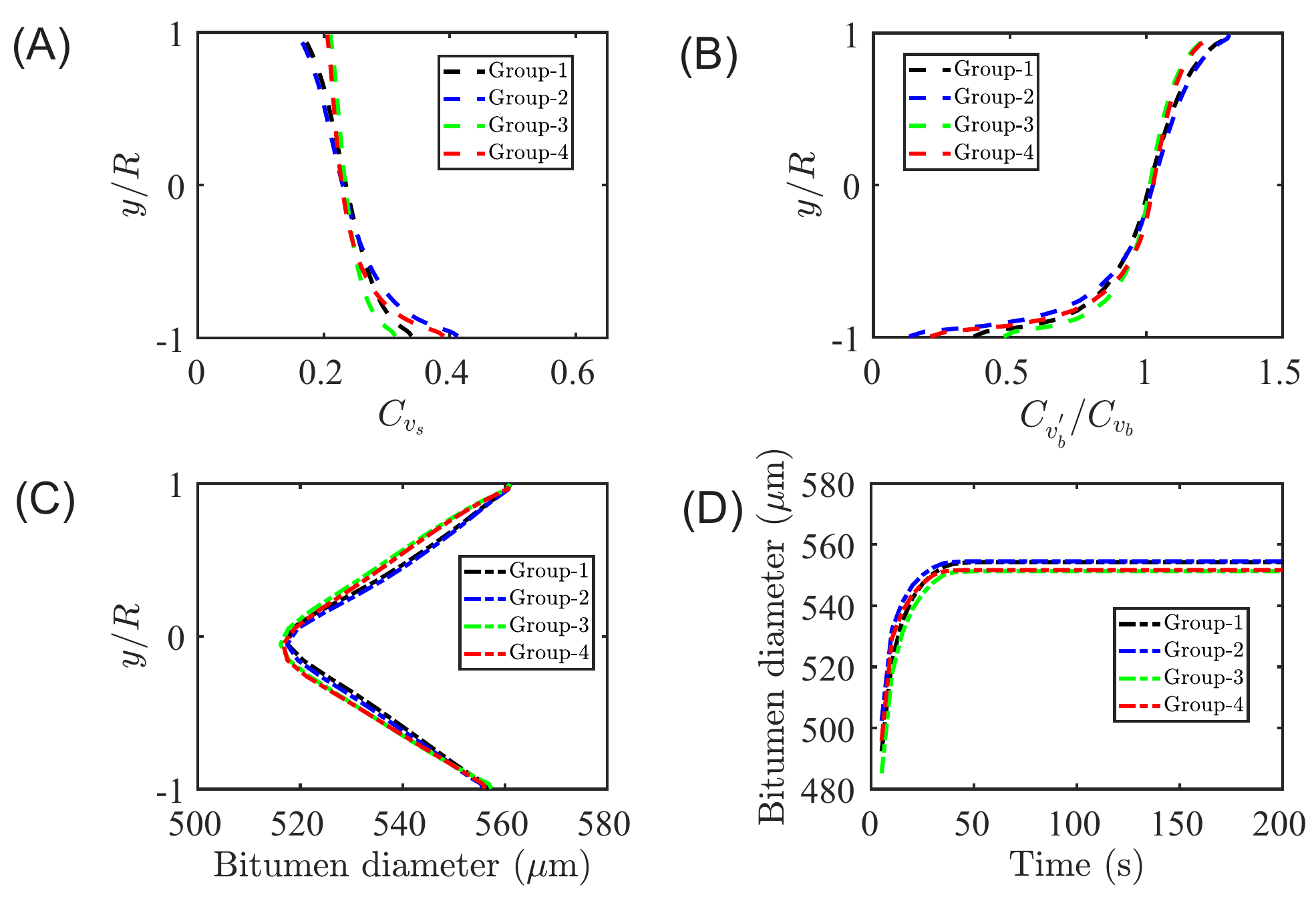}
		\caption{\label{fig:Partilce2} Effect of particle size combinations on (A) chord-average total solid concentration profiles on plane YX at Z=100 m, (B) bitumen distribution profiles along the vertical reference line at Z-100 m, (C) bitumen droplet diameter profile along the vertical reference line at Z-100 m, and (D) bitumen droplet diameter evolution with flow time on a reference point at Z= 100 m and Y= 0.30 m for a mixture velocity of $V_{m}$= 5.62 m/s.}
	\end{figure}
	
	\noindent The chord-average total solid concentration profiles are obtained for each particle size combination. Results indicate that group-1 and group-2 particles exhibit similar solid concentration profiles, with the maximum concentration at the bottom and the lowest at the top of the pipe. However, for group-4 the solid concentration is higher around the top wall of the pipe, and the maximum solid concentration occurs near the bottom wall as depicted in Fig.\ref{fig:Partilce2}A. For all the particle combinations, bitumen accumulates at the top of the pipe, and the distribution profile showed a similar curve (Fig.\ref{fig:Partilce2}B) but with a slightly higher concentration near the top wall for group-4. Bitumen droplet diameter profile along the vertical reference line show that, for all four groups, the bitumen droplet diameter is similar profile but has an apparent change for the group-3 and group-4 combinations (Fig.\ref{fig:Partilce2}C) where the s1 particle size is smaller. The evolution of bitumen droplet diameter with flow time reveals that the droplet diameter increases steadily in the initial stages and then reaches a steady state. The droplet diameter is larger for group-1 and group-2 particles compared to group-3 and group-4 particles, which have smaller droplet diameters as shown in Fig.\ref{fig:Partilce2}D.
	
	
	\begin{figure} [!ht]
		\centering
		\includegraphics[width=0.8\textwidth]{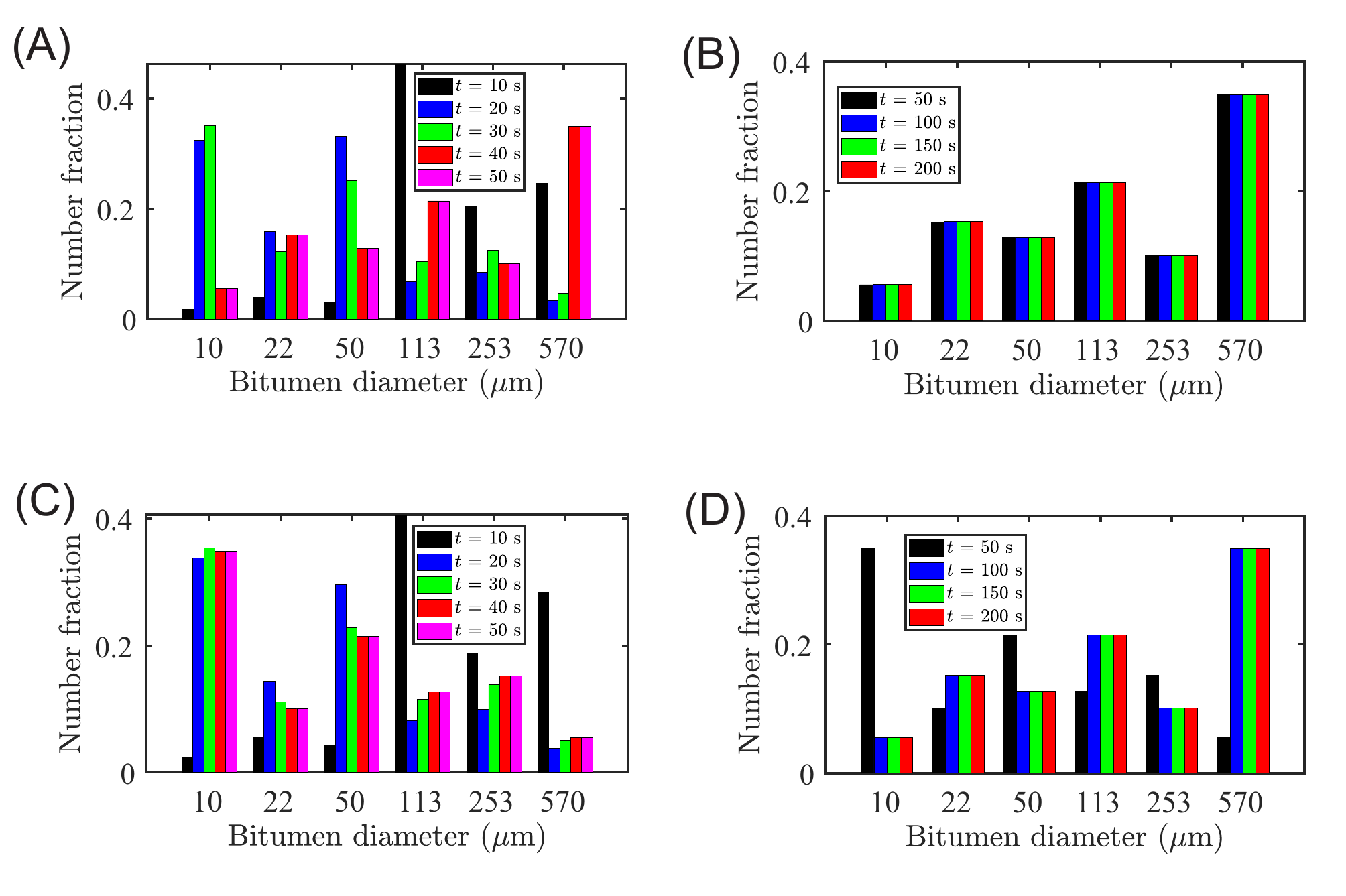}
		\caption{\label{fig:Partilce4} Initial bitumen drop size distribution (A) from 10 to 50 sec, and (B) from 50 to 200 sec group\textendash 1: S1\textendash 200 $\mu$m and S2\textendash 500 $\mu$m. (C) From 10 to 50 sec, and (D) from 50 to 200 sec group\textendash 4: S1\textendash 75 $\mu$m and S2\textendash 700 $\mu$m.at Z=100 m for a mixture velocity of $V_{m}$= 5.62 m/s.}
	\end{figure}
	In addition, the analysis of bitumen number fraction for group-1 and group-4 combinations on the plane from 10 to 50 sec revealed noticeable changes as displayed in Fig.\ref{fig:Partilce4}A and C. Specifically, the number fraction of particles with smaller sizes increased near the top of the pipe, while larger particles accumulated at the bottom for both groups. This can be attributed to the force acting on the particles, causing larger particles to settle at the bottom. However, Fig.\ref{fig:Partilce4}B and D demonstrate the absence of changes in particle number fractions from 50 to 200 sec indicating that the particles have reached a steady state, and there is no significant change in their distribution over time.
	\section{Conclusions}
	
	In this article, we demonstrate an integrated CFD-PBM model using an Eulerian-Eulerian multi-fluid model for oil sands tailings transport in an industrial-scale horizontal pipeline. We model a three-dimensional, transient Eulerian CFD model with the kinetic theory of granular flow to investigate the bitumen distribution in a complex multiphase slurry system. We implemented the CFD-PBM coupled model with the droplet breakage and aggregation phenomena for the secondary phase of the bitumen. The developed CFD-PBM was validated with six sets of field data using velocity profile and pressure comparison, demonstrating good agreement. Furthermore, we perform simulations using the CFD-PBM model to analyze the predictive capabilities of the concentrated slurry system where bitumen droplets are distributed in the flow domain.
	
	The numerical findings from this study demonstrate that slurry velocity plays a significant role in the distribution of bitumen and solids in the pipeline. The majority of the droplets accumulate near the wall of the pipe, and the aggregate bitumen droplet size is relatively higher for lower slurry velocity. After reaching a certain flow time, the aggregate size remains constant for all considered slurry velocities. Furthermore, an increase in bitumen fraction leads to a slight shift in solid concentration profiles at the bottom of the pipe, and the aggregate bitumen size increases in the middle of the pipe. With an increase in slurry velocity, the number fraction also increases for large particles near the outlet compared to small particles. The study also reveals that an increase in bubble size results in differences in bitumen distribution profiles along the length, but minimal effects on bitumen diameter profiles. Altering the particle size combination can effectively improve the distribution of bitumen and solid particles.
	
	The CFD-PBM findings revealed that lower slurry velocities tend to yield larger aggregate bitumen droplets, while higher velocities result in a more uniform distribution. Additionally, changes in bitumen fraction and bubble size were considered, highlighting their influence on the distribution profiles. These findings collectively serve as valuable criteria for assessing breakage and aggregation phenomena in our CFD-PBM model. 
	The CFD-PBM model's discrete particle size range may not capture real-world diversity. It performs well in defined flow conditions, but validation is crucial for extreme or unknown scenarios. Enhancements could involve incorporating a broader size distribution and flow range for improved predictions.
	
	The multi-fluid CFD-PBM model, accounting for droplet breakage and aggregation, is applicable in predicting secondary phase distribution in concentrated slurry flows under varying conditions. This model holds potential for industrial-scale slurry transport, particularly in the recovery of bitumen from tailings slurries. 

	\cleardoublepage
	\section*{Credit author statement}
	
	\noindent \textbf{S.G. Sontti:} Conceptualization (lead); Methodology (lead); Planned and performed the simulations (lead); Model validation (lead); Formal analysis (lead); software (lead); Visualizations (lead); Writing--original draft (lead); Writing--review \& editing (lead).\textbf{X. Zhang:} Conceptualization (supporting); Methodology (supporting); Data interpretation (lead); Project administration(lead); Writing--review \& editing (lead); Resources (lead); Supervision (lead).

	\section*{SupportingInformation}
	Computational model solver settings  (Table S1); Mesh structures along the length of the computational domain and  Cross-sectional view of the mesh structure (Figure S1); Comparison of CFD and CFD-PBM results with a constant bitumen droplet size (Table S2); Sensitivity analysis on number of bins (Table S3 \& Figure S2) are available as Supporting Information. 
	
	\section*{Declaration of competing interest}
	\noindent The authors declare that they have no known competing financial interests or personal relationships that could have appeared to influence the work reported in this paper.
	
	\section*{Acknowledgement}
	\noindent 
	This work was supported by Imperial Oil Limited and Alberta Innovates through the Institute for Oil Sands Innovation at the University of Alberta (IOSI)(Project IOSI 2019\textendash 04 (TA)), and Natural Sciences and Engineering Research Council of Canada (NSERC). This research was undertaken, in part, thanks to funding from the Canada Research Chairs Program. We also thank Compute Canada (www.computecanada.ca) for continued support through extensive access to the Compute Canada HPC Cedar and Graham clusters. 
	
	\section*{Data availability}
	\noindent The data supporting this study's findings are available from the corresponding author upon reasonable request.

	\clearpage
	\section*{Nomenclature}
	
	\begin{longtable}{ l l}
		
		\noindent $D$ &  pipe diameter (L) \\
		$R$ & pipe radius (L)\\
		$d_p$ & particle diameter (L)\\
		$C_{v}$ & chord\textendash averaged concentration (--)\\
		$g$ & gravitational acceleration (L T$^{-2}$)\\
		$g_0$ & radial distribution function (--)\\
		$p$ & locally\textendash averaged pressure (M L$^{-1}$ T$^{-2}$)\\
		$t$ & time (T) \\
		$v$ & velocity (L T$^{-1}$)\\
		$u_b$ & bitumen phase velocity (L T$^{-1}$)\\
		$V$ & velocity (L T$^{-1}$)\\
		$f_\mathrm{drag}$ & drag function (--)\\
		$n_i$ & number size distribution for a given volume of bubble size, location, and time\\
		$S_i$& source term\\
		$C_\mathrm{fr}$ & friction coefficient between solid phases (--)\\
		$B_{a,i}$ & birth term due to aggregation for the bin of index ‘i’ (number of droplets/time)\\ 
		$D_{a,i}$ & death term due to aggregation for the bin of index ‘i’ (number of droplets/time)\\
		$B_{b,i}$ &  birth term due to breakage for the bin of index ‘i’ (number of droplets/time) \\
		$D_{b,i}$& death term due to breakage for the bin of index ‘i’ (num- ber of droplets/time)\\
		$\theta$ & Aggregation kernel\\
		$\beta$ & Aggregation kernel function\\
		$ \nu  $ & volume of droplet ($L^{3}$) \\
		$n_i$& Number density of particles with size $v_i$\\
		$n_j$ & Number density of particles with size $v_j$\\
		$x$ & horizontal coordinate (L)\\
		$y$ & vertical coordinate (L)\\
		$z$ & axial coordinate (L)\\
		$e$ & restitution coefficient (--)\\
		$I_{2D}$ & second invariant of the deviatoric stress tensor (--)\\
		$\norm{\vec{v}_s^{\,\prime}}$ & fluctuating solids velocity (L T$^{-1}$)\\
		$i$ & hydraulic gradient (--)\\
		$K_{ls}$ & momentum exchange coefficient between fluid\\
		$\Delta P$ & area\textendash averaged gauge pressure (M L$^{-1}$ T$^{-2}$))\\
		$k$& turbulent kinetic energy (L$^{2}$ T$^{-2}$)) \\ 	
		\textit{Greek symbol}\\ 	
		$\alpha$ & locally\textendash averaged volume fraction (--)\\
		$\mu$ & dynamic viscosity (M L$^{-1}$ T$^{-1}$)\\
		$\rho$ & density (M L$^{-3}$)\\
		$\phi_{ls}$ & the energy exchange between the fluid and the solid phases (E) \\
		$\gamma_{\Theta_{s}}$ & collisional dissipation of energy (E)\\
		$\tau$ & shear stress (M L$^{-1}$ T$^{-2}$)\\
		$\dot{\gamma}$ & shear strain rate (T$^{-1}$)\\
		$\alpha_{s,\mathrm{max}}$ & maximum packing limit (--)\\
		$\Theta$  & granular temperature (L$^{-2}$ T$^{-2}$)\\
		$\varphi$ & angle of internal friction (--)\\
		$\eta_{t}$ & turbulent diffusivity (--)\\
		$\eta$ & apparent viscosity (M L$^{-1}$ T$^{-1}$)\\ \\
		
		\textit{Subscripts}
		\vspace{0.5cm}
		$l$ & liquid\\
		$s$ & solid\\
		$ss$ & solid particles \\
		$p$ & $p^{th}$ solid phase\\
		$q$  & $q^{th}$ solid phase\\
		col & collisional part of viscosity\\
		kin & kinetic part of viscosity\\
		fr & frictional part of viscosity  \\ 
		
	\end{longtable}
	
	\bibliography{mybibfile}

\begin{thebibliography}{50}
\expandafter\ifx\csname natexlab\endcsname\relax\def\natexlab#1{#1}\fi
\providecommand{\url}[1]{\texttt{#1}}
\providecommand{\href}[2]{#2}
\providecommand{\path}[1]{#1}
\providecommand{\DOIprefix}{doi:}
\providecommand{\ArXivprefix}{arXiv:}
\providecommand{\URLprefix}{URL: }
\providecommand{\Pubmedprefix}{pmid:}
\providecommand{\doi}[1]{\href{http://dx.doi.org/#1}{\path{#1}}}
\providecommand{\Pubmed}[1]{\href{pmid:#1}{\path{#1}}}
\providecommand{\bibinfo}[2]{#2}
\ifx\xfnm\relax \def\xfnm[#1]{\unskip,\space#1}\fi
\bibitem[{Pullum et~al.(2018)Pullum, Boger, and Sofra}]{pullum2018hydraulic}
\bibinfo{author}{L.~Pullum}, \bibinfo{author}{D.~V. Boger},
  \bibinfo{author}{F.~Sofra},
\newblock \bibinfo{title}{Hydraulic mineral waste transport and storage},
\newblock \bibinfo{journal}{Annu. Rev. Fluid Mech.} \bibinfo{volume}{58}
  (\bibinfo{year}{2018}) \bibinfo{pages}{157--185}.
\bibitem[{Plumlee and Morman(2011)}]{plumlee2011mine}
\bibinfo{author}{G.~S. Plumlee}, \bibinfo{author}{S.~A. Morman},
\newblock \bibinfo{title}{Mine wastes and human health},
\newblock \bibinfo{journal}{Elements} \bibinfo{volume}{7}
  (\bibinfo{year}{2011}) \bibinfo{pages}{399--404}.
\bibitem[{Mohaibes and Heinonen-Tanski(2004)}]{mohaibes2004aerobic}
\bibinfo{author}{M.~Mohaibes}, \bibinfo{author}{H.~Heinonen-Tanski},
\newblock \bibinfo{title}{Aerobic thermophilic treatment of farm slurry and
  food wastes},
\newblock \bibinfo{journal}{Bioresour. Technol.} \bibinfo{volume}{95}
  (\bibinfo{year}{2004}) \bibinfo{pages}{245--254}.
\bibitem[{Scoble et~al.(2003)Scoble, Klein, and Dunbar}]{scoble2003mining}
\bibinfo{author}{M.~Scoble}, \bibinfo{author}{B.~Klein}, \bibinfo{author}{W.~S.
  Dunbar},
\newblock \bibinfo{title}{Mining waste: Transforming mining systems for waste
  management},
\newblock \bibinfo{journal}{Int. J. Min. Reclam. Environ.} \bibinfo{volume}{17}
  (\bibinfo{year}{2003}) \bibinfo{pages}{123--135}.
\bibitem[{Khademi et~al.(2018)Khademi, Abbaspour, Mart{\'\i}nez-Mart{\'\i}nez,
  Gabarr{\'o}n, Shahrokh, Faz, and Acosta}]{khademi2018provenance}
\bibinfo{author}{H.~Khademi}, \bibinfo{author}{A.~Abbaspour},
  \bibinfo{author}{S.~Mart{\'\i}nez-Mart{\'\i}nez},
  \bibinfo{author}{M.~Gabarr{\'o}n}, \bibinfo{author}{V.~Shahrokh},
  \bibinfo{author}{A.~Faz}, \bibinfo{author}{J.~A. Acosta},
\newblock \bibinfo{title}{Provenance and environmental risk of windblown
  materials from mine tailing ponds, murcia, spain},
\newblock \bibinfo{journal}{Environ. Pollut.} \bibinfo{volume}{241}
  (\bibinfo{year}{2018}) \bibinfo{pages}{432--440}.
\bibitem[{Dibike et~al.(2018)Dibike, Shakibaeinia, Droppo, and
  Caron}]{dibike2018modelling}
\bibinfo{author}{Y.~B. Dibike}, \bibinfo{author}{A.~Shakibaeinia},
  \bibinfo{author}{I.~G. Droppo}, \bibinfo{author}{E.~Caron},
\newblock \bibinfo{title}{Modelling the potential effects of oil-sands tailings
  pond breach on the water and sediment quality of the lower athabasca river},
\newblock \bibinfo{journal}{Sci. Total Environ.} \bibinfo{volume}{642}
  (\bibinfo{year}{2018}) \bibinfo{pages}{1263--1281}.
\bibitem[{Ruan et~al.(2021)Ruan, Wu, B{\"u}rger, Betancourt, Wang, Wang, Jiao,
  and Wang}]{ruan2021effect}
\bibinfo{author}{Z.~Ruan}, \bibinfo{author}{A.~Wu},
  \bibinfo{author}{R.~B{\"u}rger}, \bibinfo{author}{F.~Betancourt},
  \bibinfo{author}{Y.~Wang}, \bibinfo{author}{Y.~Wang},
  \bibinfo{author}{H.~Jiao}, \bibinfo{author}{S.~Wang},
\newblock \bibinfo{title}{Effect of interparticle interactions on the yield
  stress of thickened flocculated copper mineral tailings slurry},
\newblock \bibinfo{journal}{Powder Technol.} \bibinfo{volume}{392}
  (\bibinfo{year}{2021}) \bibinfo{pages}{278--285}.
\bibitem[{Zhang et~al.(2011)Zhang, Kumar, and Scales}]{scales2011}
\bibinfo{author}{X.~Zhang}, \bibinfo{author}{A.~Kumar}, \bibinfo{author}{P.~J.
  Scales},
\newblock \bibinfo{title}{Effects of solvency and interfacial nanobubbles on
  surface forces and bubble attachment at solid surfaces},
\newblock \bibinfo{journal}{Langmuir} \bibinfo{volume}{27}
  (\bibinfo{year}{2011}) \bibinfo{pages}{2484--2491}.
\bibitem[{Zhang and Lohse(2014)}]{zhang2014}
\bibinfo{author}{X.~Zhang}, \bibinfo{author}{D.~Lohse},
\newblock \bibinfo{title}{{Perspectives on surface nanobubbles}},
\newblock \bibinfo{journal}{Biomicrofluidics} \bibinfo{volume}{8}
  (\bibinfo{year}{2014}).
\bibitem[{Motamed~Dashliborun et~al.(2020)Motamed~Dashliborun, Zhou, Esmaeili,
  and Zhang}]{motamed2020microbubble}
\bibinfo{author}{A.~Motamed~Dashliborun}, \bibinfo{author}{J.~Zhou},
  \bibinfo{author}{P.~Esmaeili}, \bibinfo{author}{X.~Zhang},
\newblock \bibinfo{title}{Microbubble-enhanced recovery of residual bitumen
  from the tailings of oil sands extraction in a laboratory-scale pipeline},
\newblock \bibinfo{journal}{Energy \& Fuels} \bibinfo{volume}{34}
  (\bibinfo{year}{2020}) \bibinfo{pages}{16476--16485}.
\bibitem[{Zhou et~al.(2022)Zhou, Sontti, Zhou, Esmaeili, and
  Zhang}]{zhou2022microbubble}
\bibinfo{author}{K.~Zhou}, \bibinfo{author}{S.~G. Sontti},
  \bibinfo{author}{J.~Zhou}, \bibinfo{author}{P.~Esmaeili},
  \bibinfo{author}{X.~Zhang},
\newblock \bibinfo{title}{Microbubble-enhanced bitumen separation from tailing
  slurries with high solid contents},
\newblock \bibinfo{journal}{Ind. Eng. Chem. Res.} \bibinfo{volume}{61}
  (\bibinfo{year}{2022}) \bibinfo{pages}{17327--17341}.
\bibitem[{Zhou et~al.(2023)Zhou, Sontti, Zhou, and Zhang}]{zhou2023}
\bibinfo{author}{K.~Zhou}, \bibinfo{author}{S.~G. Sontti},
  \bibinfo{author}{J.~Zhou}, \bibinfo{author}{X.~Zhang},
\newblock \bibinfo{title}{Enhanced accumulation of bitumen residue in a highly
  concentrated tailings flow by microbubbles from in-situ catalytic
  decomposition of hydrogen peroxide},
\newblock \bibinfo{journal}{Fuel} \bibinfo{volume}{345} (\bibinfo{year}{2023})
  \bibinfo{pages}{128249}.
\bibitem[{Sadeghi et~al.(2023)Sadeghi, Sontti, Zheng, and
  Zhang}]{sadeghi2023computational}
\bibinfo{author}{M.~Sadeghi}, \bibinfo{author}{S.~G. Sontti},
  \bibinfo{author}{E.~Zheng}, \bibinfo{author}{X.~Zhang},
\newblock \bibinfo{title}{Computational fluid dynamics (cfd) simulation of
  three--phase non--newtonian slurry flows in industrial horizontal pipelines},
\newblock \bibinfo{journal}{Chem. Eng. Sci.}  (\bibinfo{year}{2023})
  \bibinfo{pages}{118513}.
\bibitem[{Messa et~al.(2021)Messa, Yang, Adedeji, Ch{\'a}ra, Duarte,
  Matou{\v{s}}ek, Rasteiro, Sanders, Silva, and
  de~Souza}]{messa2021computational}
\bibinfo{author}{G.~V. Messa}, \bibinfo{author}{Q.~Yang},
  \bibinfo{author}{O.~E. Adedeji}, \bibinfo{author}{Z.~Ch{\'a}ra},
  \bibinfo{author}{C.~A.~R. Duarte}, \bibinfo{author}{V.~Matou{\v{s}}ek},
  \bibinfo{author}{M.~G. Rasteiro}, \bibinfo{author}{R.~S. Sanders},
  \bibinfo{author}{R.~C. Silva}, \bibinfo{author}{F.~J. de~Souza},
\newblock \bibinfo{title}{Computational fluid dynamics modelling of
  liquid--solid slurry flows in pipelines: State-of-the-art and future
  perspectives},
\newblock \bibinfo{journal}{Processes} \bibinfo{volume}{9}
  (\bibinfo{year}{2021}) \bibinfo{pages}{1566}.
\bibitem[{Parvathaneni and Buwa(2021)}]{parvathaneni2021eulerian}
\bibinfo{author}{S.~Parvathaneni}, \bibinfo{author}{V.~V. Buwa},
\newblock \bibinfo{title}{Eulerian multifluid simulations of segregation and
  mixing of binary gas-solids flow of particles with different densities},
\newblock \bibinfo{journal}{Chem. Eng. Sci.} \bibinfo{volume}{245}
  (\bibinfo{year}{2021}) \bibinfo{pages}{116901}.
\bibitem[{Sontti et~al.(2023)Sontti, Sadeghi, Zhou, Zheng, and
  Zhang}]{sontti2022computational}
\bibinfo{author}{S.~G. Sontti}, \bibinfo{author}{M.~Sadeghi},
  \bibinfo{author}{K.~Zhou}, \bibinfo{author}{E.~Zheng},
  \bibinfo{author}{X.~Zhang},
\newblock \bibinfo{title}{Computational fluid dynamics investigation of bitumen
  residues in oil sands tailings transport in an industrial horizontal pipe},
\newblock \bibinfo{journal}{Phys. Fluids} \bibinfo{volume}{35}
  (\bibinfo{year}{2023}).
\bibitem[{Cai et~al.(2019)Cai, Liu, Mi, Luo, Ma, Xu, and
  Yang}]{cai2019investigation}
\bibinfo{author}{L.~Cai}, \bibinfo{author}{Z.~Liu}, \bibinfo{author}{S.~Mi},
  \bibinfo{author}{C.~Luo}, \bibinfo{author}{K.~Ma}, \bibinfo{author}{A.~Xu},
  \bibinfo{author}{S.~Yang},
\newblock \bibinfo{title}{Investigation on flow characteristics of ice slurry
  in horizontal 90° elbow pipe by a cfd-pbm coupled model},
\newblock \bibinfo{journal}{Adv. Powder Technol.} \bibinfo{volume}{30}
  (\bibinfo{year}{2019}) \bibinfo{pages}{2299--2310}.
\bibitem[{Ma et~al.(2022)Ma, Liu, Tang, Liu, Yang, and Yang}]{ma2022numerical}
\bibinfo{author}{K.~Ma}, \bibinfo{author}{Z.~Liu}, \bibinfo{author}{Y.~Tang},
  \bibinfo{author}{X.~Liu}, \bibinfo{author}{Y.~Yang},
  \bibinfo{author}{S.~Yang},
\newblock \bibinfo{title}{Numerical investigation on ice slurry flow in
  horizontal elbow pipes},
\newblock \bibinfo{journal}{Therm. Sci. Eng. Prog.} \bibinfo{volume}{27}
  (\bibinfo{year}{2022}) \bibinfo{pages}{101083}.
\bibitem[{Cao et~al.(2022)Cao, Yang, Wang, and Bian}]{cao2022gas}
\bibinfo{author}{X.~Cao}, \bibinfo{author}{K.~Yang}, \bibinfo{author}{H.~Wang},
  \bibinfo{author}{J.~Bian},
\newblock \bibinfo{title}{Gas--liquid--hydrate flow characteristics in vertical
  pipe considering bubble and particle coalescence and breakage},
\newblock \bibinfo{journal}{Chem. Eng. Sci.} \bibinfo{volume}{252}
  (\bibinfo{year}{2022}) \bibinfo{pages}{117249}.
\bibitem[{Balakin et~al.(2016)Balakin, Lo, Kosinski, and
  Hoffmann}]{balakin2016modelling}
\bibinfo{author}{B.~V. Balakin}, \bibinfo{author}{S.~Lo},
  \bibinfo{author}{P.~Kosinski}, \bibinfo{author}{A.~C. Hoffmann},
\newblock \bibinfo{title}{Modelling agglomeration and deposition of gas
  hydrates in industrial pipelines with combined cfd-pbm technique},
\newblock \bibinfo{journal}{Chem. Eng. Sci.} \bibinfo{volume}{153}
  (\bibinfo{year}{2016}) \bibinfo{pages}{45--57}.
\bibitem[{Bulmer and Starr(1979)}]{bulmer1979syncrude}
\bibinfo{author}{J.~Bulmer}, \bibinfo{author}{J.~Starr},
\newblock \bibinfo{title}{Syncrude analytical methods for oil sand and bitumen
  processing, research organizations:syncrude canada ltd., edmonton, ab
  (canada),isbn 0-7732-0852-6; trn: Ca9600967, 186}  (\bibinfo{year}{1979})
  \bibinfo{pages}{1--186}.
\bibitem[{Maron et~al.(2008)Maron, Fernald, O’Keefe, Viega, and
  Bailey}]{maron2008new}
\bibinfo{author}{R.~J. Maron}, \bibinfo{author}{M.~Fernald},
  \bibinfo{author}{C.~O’Keefe}, \bibinfo{author}{J.~Viega},
  \bibinfo{author}{T.~Bailey},
\newblock \bibinfo{title}{New applications of sonar-based technology in the
  minerals processing industry: Velocity profile measurement and pipe wall wear
  monitoring in hydrotransport lines},
\newblock \bibinfo{journal}{Technical Paper No. BI0351}  (\bibinfo{year}{2008})
  \bibinfo{pages}{1--11}.
\bibitem[{Parvathaneni et~al.(2023{\natexlab{a}})Parvathaneni, Karmakar, and
  Buwa}]{parvathaneni2023eulerian}
\bibinfo{author}{S.~Parvathaneni}, \bibinfo{author}{S.~Karmakar},
  \bibinfo{author}{V.~V. Buwa},
\newblock \bibinfo{title}{Eulerian simulations of bubbling and jetting regimes
  in a fluidized bed},
\newblock \bibinfo{journal}{Particuology} \bibinfo{volume}{75}
  (\bibinfo{year}{2023}{\natexlab{a}}) \bibinfo{pages}{50--68}.
\bibitem[{Parvathaneni et~al.(2023{\natexlab{b}})Parvathaneni, Karmakar, and
  Buwa}]{parvathaneni2023effect}
\bibinfo{author}{S.~Parvathaneni}, \bibinfo{author}{S.~Karmakar},
  \bibinfo{author}{V.~V. Buwa},
\newblock \bibinfo{title}{Effect of local hydrodynamics on the performance of a
  fluidized-bed gasifier},
\newblock \bibinfo{journal}{Ind. Eng. Chem. Res.} \bibinfo{volume}{62}
  (\bibinfo{year}{2023}{\natexlab{b}}) \bibinfo{pages}{11814--11830}.
\bibitem[{Messa and Matou{\v{s}}ek(2020)}]{messa2020analysis}
\bibinfo{author}{G.~V. Messa}, \bibinfo{author}{V.~Matou{\v{s}}ek},
\newblock \bibinfo{title}{Analysis and discussion of two fluid modelling of
  pipe flow of fully suspended slurry},
\newblock \bibinfo{journal}{Powder Technol.} \bibinfo{volume}{360}
  (\bibinfo{year}{2020}) \bibinfo{pages}{747--768}.
\bibitem[{Wang et~al.(2023)Wang, Hu, Luo, Yu, and Fan}]{wang2022multi}
\bibinfo{author}{S.~Wang}, \bibinfo{author}{C.~Hu}, \bibinfo{author}{K.~Luo},
  \bibinfo{author}{J.~Yu}, \bibinfo{author}{J.~Fan},
\newblock \bibinfo{title}{Multi-scale numerical simulation of fluidized beds:
  Model applicability assessment},
\newblock \bibinfo{journal}{Particuology} \bibinfo{volume}{80}
  (\bibinfo{year}{2023}) \bibinfo{pages}{11--41}.
\bibitem[{Alobaid et~al.(2022)Alobaid, Almohammed, {Massoudi Farid}, May,
  Rößger, Richter, and Epple}]{ALOBAID2022}
\bibinfo{author}{F.~Alobaid}, \bibinfo{author}{N.~Almohammed},
  \bibinfo{author}{M.~{Massoudi Farid}}, \bibinfo{author}{J.~May},
  \bibinfo{author}{P.~Rößger}, \bibinfo{author}{A.~Richter},
  \bibinfo{author}{B.~Epple},
\newblock \bibinfo{title}{Progress in cfd simulations of fluidized beds for
  chemical and energy process engineering},
\newblock \bibinfo{journal}{Prog. Energy Combust. Sci.} \bibinfo{volume}{91}
  (\bibinfo{year}{2022}) \bibinfo{pages}{100930}.
\bibitem[{Adeyinka et~al.(2009)Adeyinka, Samiei, Xu, and Masliyah}]{rheology}
\bibinfo{author}{O.~B. Adeyinka}, \bibinfo{author}{S.~Samiei},
  \bibinfo{author}{Z.~Xu}, \bibinfo{author}{J.~H. Masliyah},
\newblock \bibinfo{title}{Effect of particle size on the rheology of athabasca
  clay suspensions},
\newblock \bibinfo{journal}{Can. J. Chem. Eng.} \bibinfo{volume}{87}
  (\bibinfo{year}{2009}) \bibinfo{pages}{422--434}.
\bibitem[{Ansys(2011)}]{fluent2011ansys}
\bibinfo{author}{Ansys},
\newblock \bibinfo{title}{Ansys fluent theory guide},
\newblock \bibinfo{journal}{Ansys Inc., USA}  (\bibinfo{year}{2011}).
\bibitem[{Sadeghi et~al.(2022)Sadeghi, Li, Zheng, Sontti, Esmaeili, and
  Zhang}]{sadeghi2022cfd}
\bibinfo{author}{M.~Sadeghi}, \bibinfo{author}{S.~Li},
  \bibinfo{author}{E.~Zheng}, \bibinfo{author}{S.~G. Sontti},
  \bibinfo{author}{P.~Esmaeili}, \bibinfo{author}{X.~Zhang},
\newblock \bibinfo{title}{Cfd simulation of turbulent non-newtonian slurry
  flows in horizontal pipelines},
\newblock \bibinfo{journal}{Ind. Eng. Chem. Res.} \bibinfo{volume}{61}
  (\bibinfo{year}{2022}) \bibinfo{pages}{5324--5339}.
\bibitem[{Gidaspow et~al.(1991)Gidaspow, Bezburuah, and Ding}]{gidaspow1991hyd}
\bibinfo{author}{D.~Gidaspow}, \bibinfo{author}{R.~Bezburuah},
  \bibinfo{author}{J.~Ding}, \bibinfo{title}{Hydrodynamics of circulating
  fluidized beds: kinetic theory approach , Technical Report No. CONF-920502-1
  and ON: DE92002879 (Illinois Institute of Technology, Chicago, IL, 8)},
  \bibinfo{type}{Technical Report}, Illinois Inst. of Tech., Chicago, IL
  (United States). Dept. of Chemical, \bibinfo{year}{1991}.
\bibitem[{Sen et~al.(2016)Sen, Singh, Patwardhan, Mukhopadhyay, and
  Shenoy}]{sen2016cfd}
\bibinfo{author}{N.~Sen}, \bibinfo{author}{K.~Singh},
  \bibinfo{author}{A.~Patwardhan}, \bibinfo{author}{S.~Mukhopadhyay},
  \bibinfo{author}{K.~Shenoy},
\newblock \bibinfo{title}{Cfd simulation of two-phase flow in pulsed
  sieve-plate column--identification of a suitable drag model to predict
  dispersed phase hold up},
\newblock \bibinfo{journal}{Sep. Sci. Technol.} \bibinfo{volume}{51}
  (\bibinfo{year}{2016}) \bibinfo{pages}{2790--2803}.
\bibitem[{Li et~al.(2018)Li, He, Liu, and Huang}]{li2018hydrodynamic}
\bibinfo{author}{M.-z. Li}, \bibinfo{author}{Y.-p. He}, \bibinfo{author}{Y.-d.
  Liu}, \bibinfo{author}{C.~Huang},
\newblock \bibinfo{title}{Hydrodynamic simulation of multi-sized high
  concentration slurry transport in pipelines},
\newblock \bibinfo{journal}{Ocean Eng.} \bibinfo{volume}{163}
  (\bibinfo{year}{2018}) \bibinfo{pages}{691--705}.
\bibitem[{Liu et~al.(2022)Liu, He, Li, Huang, and Liu}]{liu2022effect}
\bibinfo{author}{W.~Liu}, \bibinfo{author}{Y.~He}, \bibinfo{author}{M.~Li},
  \bibinfo{author}{C.~Huang}, \bibinfo{author}{Y.~Liu},
\newblock \bibinfo{title}{Effect of drag models on hydrodynamic behaviors of
  slurry flows in horizontal pipes},
\newblock \bibinfo{journal}{Phys. Fluids} \bibinfo{volume}{34}
  (\bibinfo{year}{2022}) \bibinfo{pages}{103311}.
\bibitem[{Puhan et~al.(2023)Puhan, Mukherjee, and Atta}]{puhan2023insights}
\bibinfo{author}{P.~Puhan}, \bibinfo{author}{A.~K. Mukherjee},
  \bibinfo{author}{A.~Atta},
\newblock \bibinfo{title}{Insights into the influence of particle density and
  column inclination in polydisperse liquid--solid fluidized beds},
\newblock \bibinfo{journal}{Powder Technol.}  (\bibinfo{year}{2023})
  \bibinfo{pages}{118540}.
\bibitem[{Li et~al.(2018)Li, He, Liu, and Huang}]{li2018effect}
\bibinfo{author}{M.~Li}, \bibinfo{author}{Y.~He}, \bibinfo{author}{Y.~Liu},
  \bibinfo{author}{C.~Huang},
\newblock \bibinfo{title}{Effect of interaction of particles with different
  sizes on particle kinetics in multi-sized slurry transport by pipeline},
\newblock \bibinfo{journal}{Powder Technol.} \bibinfo{volume}{338}
  (\bibinfo{year}{2018}) \bibinfo{pages}{915--930}.
\bibitem[{Shi et~al.(2020)Shi, Li, Liu, and Nikrityuk}]{shi2020experimental}
\bibinfo{author}{H.~Shi}, \bibinfo{author}{M.~Li}, \bibinfo{author}{Q.~Liu},
  \bibinfo{author}{P.~Nikrityuk},
\newblock \bibinfo{title}{Experimental and numerical study of cavitating
  particulate flows in a venturi tube},
\newblock \bibinfo{journal}{Chem. Eng. Sci.} \bibinfo{volume}{219}
  (\bibinfo{year}{2020}) \bibinfo{pages}{115598}.
\bibitem[{Brito et~al.(2021)Brito, Jerez, and Gutierrez}]{britocasson}
\bibinfo{author}{G.~Brito}, \bibinfo{author}{O.~Jerez},
  \bibinfo{author}{L.~Gutierrez},
\newblock \bibinfo{title}{Incorporation of rheological characterization in
  grinding and tailings slurries to optimize the cmp magnetic separation
  plant},
\newblock \bibinfo{journal}{Minerals} \bibinfo{volume}{11}
  (\bibinfo{year}{2021}).
\bibitem[{Ramkrishna(2000)}]{ramkrishna2000population}
\bibinfo{author}{D.~Ramkrishna}, \bibinfo{title}{Population balances: Theory
  and applications to particulate systems in engineering},
  \bibinfo{publisher}{Elsevier}, \bibinfo{year}{2000}.
\bibitem[{Luo and Svendsen(1996)}]{luo1996theoretical}
\bibinfo{author}{H.~Luo}, \bibinfo{author}{H.~F. Svendsen},
\newblock \bibinfo{title}{Theoretical model for drop and bubble breakup in
  turbulent dispersions},
\newblock \bibinfo{journal}{AIChE J.} \bibinfo{volume}{42}
  (\bibinfo{year}{1996}) \bibinfo{pages}{1225--1233}.
\bibitem[{Li et~al.(2021)Li, Lian, Bai, Zhao, Li, Zhang, and Huang}]{li2021cfd}
\bibinfo{author}{L.~Li}, \bibinfo{author}{W.~Lian}, \bibinfo{author}{B.~Bai},
  \bibinfo{author}{Y.~Zhao}, \bibinfo{author}{P.~Li},
  \bibinfo{author}{Q.~Zhang}, \bibinfo{author}{W.~Huang},
\newblock \bibinfo{title}{Cfd-pbm investigation of the hydrodynamics in a
  slurry bubble column reactor with a circular gas distributor and heat
  exchanger tube},
\newblock \bibinfo{journal}{Chem. Eng. Sci.: X} \bibinfo{volume}{9}
  (\bibinfo{year}{2021}) \bibinfo{pages}{100087}.
\bibitem[{Thakur et~al.(2022)Thakur, Kumar, Banerjee, Chaudhari, and
  Gaurav}]{thakur2022hydrodynamic}
\bibinfo{author}{A.~K. Thakur}, \bibinfo{author}{R.~Kumar},
  \bibinfo{author}{N.~Banerjee}, \bibinfo{author}{P.~Chaudhari},
  \bibinfo{author}{G.~K. Gaurav},
\newblock \bibinfo{title}{Hydrodynamic modeling of liquid-solid flow in
  polyolefin slurry reactors using cfd techniques--a critical analysis},
\newblock \bibinfo{journal}{Powder Technol.}  (\bibinfo{year}{2022})
  \bibinfo{pages}{117544}.
\bibitem[{Maluta et~al.(2023)Maluta, Paglianti, and
  Montante}]{maluta2023experimental}
\bibinfo{author}{F.~Maluta}, \bibinfo{author}{A.~Paglianti},
  \bibinfo{author}{G.~Montante},
\newblock \bibinfo{title}{Experimental and numerical study of a compact inline
  swirler for gas--liquid separation},
\newblock \bibinfo{journal}{Chem. Eng. Sci.} \bibinfo{volume}{265}
  (\bibinfo{year}{2023}) \bibinfo{pages}{118219}.
\bibitem[{Zhang et~al.(2021)Zhang, Nathan, Tian, and Chin}]{zhang2021influence}
\bibinfo{author}{X.~Zhang}, \bibinfo{author}{G.~J. Nathan},
  \bibinfo{author}{Z.~F. Tian}, \bibinfo{author}{R.~C. Chin},
\newblock \bibinfo{title}{The influence of the coefficient of restitution on
  flow regimes within horizontal particle-laden pipe flows},
\newblock \bibinfo{journal}{Phys. Fluids} \bibinfo{volume}{33}
  (\bibinfo{year}{2021}) \bibinfo{pages}{123318}.
\bibitem[{Liu et~al.(2021)Liu, He, Li, Chen, Liu, and
  Huang}]{liu2021computational}
\bibinfo{author}{W.~Liu}, \bibinfo{author}{Y.~He}, \bibinfo{author}{M.~Li},
  \bibinfo{author}{Q.~Chen}, \bibinfo{author}{Y.~Liu},
  \bibinfo{author}{C.~Huang},
\newblock \bibinfo{title}{Computational fluid dynamics modeling of slurry flow
  in horizontal pipes: Effect of specularity coefficient on hydraulic
  gradient},
\newblock \bibinfo{journal}{Ocean Eng.} \bibinfo{volume}{238}
  (\bibinfo{year}{2021}) \bibinfo{pages}{109625}.
\bibitem[{Zheng et~al.(2021)Zheng, Rudman, Kuang, and
  Chryss}]{zheng2021turbulent}
\bibinfo{author}{E.~Zheng}, \bibinfo{author}{M.~Rudman},
  \bibinfo{author}{S.~Kuang}, \bibinfo{author}{A.~Chryss},
\newblock \bibinfo{title}{Turbulent coarse-particle non-newtonian suspension
  flow in a pipe},
\newblock \bibinfo{journal}{Int. J. Multiph. Flow .} \bibinfo{volume}{142}
  (\bibinfo{year}{2021}) \bibinfo{pages}{103698}.
\bibitem[{Shook and Roco(2013)}]{shook2013slurry}
\bibinfo{author}{C.~A. Shook}, \bibinfo{author}{M.~C. Roco},
  \bibinfo{title}{Slurry flow: principles and practice},
  \bibinfo{publisher}{Elsevier}, \bibinfo{year}{2013}.
\bibitem[{Zahid et~al.(2020)Zahid, Ur~Rehman, Rushd, Hasan, and
  Rahman}]{zahid2020experimental}
\bibinfo{author}{A.~A. Zahid}, \bibinfo{author}{S.~R. Ur~Rehman},
  \bibinfo{author}{S.~Rushd}, \bibinfo{author}{A.~Hasan},
  \bibinfo{author}{M.~A. Rahman},
\newblock \bibinfo{title}{Experimental investigation of multiphase flow
  behavior in drilling annuli using high speed visualization technique},
\newblock \bibinfo{journal}{Front. Energy .} \bibinfo{volume}{14}
  (\bibinfo{year}{2020}) \bibinfo{pages}{635--643}.
\bibitem[{Rosas et~al.(2018)Rosas, Bassani, Alves, Schneider, Marcelino~Neto,
  Morales, and Sum}]{rosas2018measurements}
\bibinfo{author}{L.~M. Rosas}, \bibinfo{author}{C.~L. Bassani},
  \bibinfo{author}{R.~F. Alves}, \bibinfo{author}{F.~A. Schneider},
  \bibinfo{author}{M.~A. Marcelino~Neto}, \bibinfo{author}{R.~E. Morales},
  \bibinfo{author}{A.~K. Sum},
\newblock \bibinfo{title}{Measurements of horizontal three-phase
  solid-liquid-gas slug flow: Influence of hydrate-like particles on
  hydrodynamics},
\newblock \bibinfo{journal}{AIChE Journal} \bibinfo{volume}{64}
  (\bibinfo{year}{2018}) \bibinfo{pages}{2864--2880}.
\bibitem[{Parvathaneni and Buwa(2020)}]{parvathaneni2020role}
\bibinfo{author}{S.~Parvathaneni}, \bibinfo{author}{V.~V. Buwa},
\newblock \bibinfo{title}{Role of bubbling behaviour in segregation and mixing
  of binary gas-solids flow of particles with different density},
\newblock \bibinfo{journal}{Powder Technol.} \bibinfo{volume}{372}
  (\bibinfo{year}{2020}) \bibinfo{pages}{178--191}.

\end{thebibliography}
	
	\cleardoublepage
\end{document}